\documentstyle[preprint,aps]{revtex}
\begin{document}
\draft
\title{High-K Isomers in $^{176}$W and Mechanisms
of K-Violation}

\author{B. Crowell,$^{(a)}$  P. Chowdhury,$^{(b)}$  D.J. Blumenthal,$^{(a)}$
 S.J. Freeman$^{(c)}$ and C.J. Lister$^{(a)}$}
\address{Wright Nuclear Structure Laboratory,  Yale
University,  New Haven, CT 06511}

\author{M.P. Carpenter,  R.G. Henry$^{(d)}$,  R.V.F. Janssens,
T.L. Khoo,  T. Lauritsen,  Y. Liang$^{(e)}$ and
F. Soramel$^{(f)}$}
\address{Argonne National Laboratory,  Argonne, IL
60439}

\author{I.G. Bearden$^{(g)}$}
\address{Purdue University,  West Lafayette, IN 47907}

\maketitle

\vspace{0.01in}

\begin{abstract}

An isomer, with t$_{1/2}$=35 $\pm$ 10 ns and
J, K$^{\pi}$ = 14, 14$^{+}$, has been observed
in the nucleus $^{176}$W using the reaction
$^{150}$Nd($^{30}$Si,4n) at a beam energy
of 133 MeV.  The isomer exhibits an unusual
pattern of decay in which the {\em majority} of
the flux proceeds directly to states with
$<$K$>$=0, bypassing available levels of intermediate
K.  This severe breakdown of normal K-selection
rules in $^{176}$W is compared with recent
observations of K-violation in neighboring
nuclei, within the framework of proposed theoretical
approaches.  The available data on these K-violating
decays seem to have a consistent explanation
in models of K-mixing which include large-amplitude
fluctuations of the nuclear shape.


$^{(a)}$ Present address: Argonne National Laboratory,
Argonne, IL 60439.

$^{(b)}$ Present address: Dept. of Physics,  Wellesley
College,  Wellesley, MA 02181 and Dept. of Physics,
Univ. of Massachusetts, Lowell, MA 01854.

$^{(c)}$ Present address: Schuster Laboratory,
University of Manchester,  Manchester M13 9PL,
United Kingdom.

$^{(d)}$Dept. of Radiology, University of
California at San Francisco, San Francisco, CA 94143.

$^{(e)}$ Present address: Dept. of Radiology,
Indiana Univ. Medical Center,  Indianapolis,
IN 46202.

$^{(f)}$ Present address: Dept. of Physics,  University
of Udine,  I-33100 Udine, Italy, on leave from
Universit\'{a} di Padova, Italy.

$^{(g)}$ Present address: Niels Bohr Institute,
DK-4000 Roskilde, Denmark

\end{abstract}
\vspace{0.01in}
\thispagestyle{empty}

\vspace{0.01in}

\noindent
--------------- \\
\noindent

\section{INTRODUCTION}

	A quantum-mechanical system containing
a finite number of particles can
often be accurately
described in terms of a shape.  For particles
moving in a mean field imposed by such a shape,
symmetries of the shape imply the existence
of conserved quantum numbers.  In deformed
nuclei with axially symmetric equilibrium
shapes,  the projection, K, of the total angular
momentum along the axis of symmetry is an
example of such an approximately conserved
quantum number.  The yrast states  (i.e. the
states with the lowest energy for a given
angular momentum) in even-even nuclei
with such shapes
are most commonly those involving collective
rotation, with K=0.  Collective rotation
does not contribute to K,  because a quantum
rotor cannot rotate about an axis of symmetry.
High-K states compete with collective rotation
as an energy-efficient mode of accommodating
angular momentum only in a small number of
nuclei near A=180.  These nuclei have many
orbitals near the neutron and proton Fermi
surfaces with large projections, $\Omega$,
of the
individual nucleonic angular momenta along
the axis of symmetry.

	Systematic studies have revealed a normal
pattern of decay of high-K states which proceeds
stepwise to lower values of K, minimizing
$\Delta$K at each step \cite{Boh75}.  When
the only available decay mode is via a transition
with a large $\Delta$K, the decay is
hindered and the level is isomeric.  The existence
of such isomers, known as K-isomers, is a
clear indication of the approximate conservation
of the K quantum number.  The associated selection
rule involves the degree of K-forbiddenness,
$\nu$, defined as
$\nu$=$\mid$$\Delta$K$\mid$-$\lambda$,  where
$\lambda$ is the multipolarity of
the gamma-ray transition associated with the
decay.
Thus, $\nu$ denotes the part of $\Delta$K
that cannot be accounted for by the
angular momentum carried away by the photon.
Transitions with $\nu$ $>$ 0 are forbidden to first
order and are hindered.  A high-K isomer is
able to decay to a state of lower K only because
of very small admixtures of non-dominant values
of K in the wave-functions of the initial
or final state, or both.  An empirical measure
of K-mixing is the hindrance factor,  F =
t$_{1/2}$/t$^{W}_{1/2}$,  where t$_{1/2}$ is the
experimentally determined partial half-life
of the transition,  and t$^{W}_{1/2}$ is the Weisskopf
estimate for the half-life of a single-particle
transition.  The typical observation is that
F increases by a factor of about 100 per degree
of K-forbiddenness \cite{Lob68}.

	One mechanism that has clearly been shown
to contribute to K-mixing,  especially for
small values of $\Delta$K,  is Coriolis mixing
with a fixed shape \cite{Boh75}.
This represents fluctuations in the
angle between the axis of symmetry and the
angular momentum vector.  In the rotating
body-fixed frame,  the Coriolis force contributes
to the Hamiltonian a term
H$_{Cor}$ = $-\omega$J$_{x}$,
where $\omega$ is the rotational frequency, and
J$_{x}$ is the component of the angular momentum
perpendicular to the axis of symmetry.  The
matrix elements of H$_{Cor}$ link states differing
by $\Delta$K=$\pm$1, and are typically small
compared to the spacing of states with a given
K-value,  so that a perturbative treatment
of the mixing converges rapidly.  Coriolis
admixtures with large $\Delta$K occur
only in high-order perturbation theory,  and
should show a roughly exponential increase
of F as a function of $\nu$.  This is precisely
what was observed in all the cases originally
studied  \cite{Lob68},  and until recently,
the physical mechanisms of K-mixing were
thought to be fairly well understood in terms
of Coriolis couplings.

	More recently, transitions with large $\Delta$K,
but abnormally low hindrance factors, have
been observed in a few nuclei
\cite{Cho88,Hf174,Wal94a}.
This constitutes a challenge to the established
picture of Coriolis mixing.  A new mechanism that has
been suggested to explain these abnormal cases
\cite{Cho88,Ben89d} involves large-amplitude
quantum fluctuations of the shape away from
axial symmetry, measured by the triaxiality
parameter $\gamma$.  Small non-axial ellipsoidal
fluctuations cause $\Delta$K = $\pm$2 mixing,
and can therefore only cause large-$\Delta$K
admixtures through higher-order effects.
Large non-axial fluctuations,  however,  may
result in a complete rearrangement of the
nuclear wave-function,  and cannot be understood
perturbatively.  Such fluctuations could carry
a nucleus from $\gamma$=--120$^{\circ}$
(an axially symmetric
prolate shape with K=J ) to
$\gamma$=0$^{\circ}$ (the same
shape with K=0),  penetrating a barrier in
the potential at intermediate values of $\gamma$
\cite{Cho88,Ben89d}.  Such a mechanism is
referred to as $\gamma$-tunneling.
The definition of
$\gamma$ adopted here corresponds to
the Lund convention.

	Our goal in this work was to find
experimental
observations and theoretical methods that
would allow a comparison of simple models
of $\gamma$-tunneling and Coriolis mixing
as descriptions of
these anomalous decays of K-isomers.
If $\gamma$-tunneling
is an important mechanism in these anomalous
decays,  then the measured hindrance factors
should vary strongly as a function of the shape and
height of the potential barrier between the
minima at $\gamma$=--120$^{\circ}$ and
$\gamma$=0$^{\circ}$.  In the A$\sim$180
region,  the height of this barrier is predicted
to vary rapidly with changing proton number,
based on calculations of the potential energy
surfaces described in section IV.
In agreement with these calculations, the
measured excitation energies of
$\gamma$-vibrational
bands also show significant changes \cite{Soo91},
reflecting variations in the softness
towards triaxial deformations.  If Coriolis
mixing is the dominant mechanism,  then the
strengths of the decays should be a function
of variables such as the quasiparticle  (qp)
configuration,
the deformation,  and the rotational frequency,
and should not depend on the shape of the
potential energy surface.

	The nucleus $^{176}$W is a good candidate
for testing these models.  It is an isotone
of $^{174}$Hf,  in which weak branches with
large $\Delta$K have been observed \cite{Hf174}
in the decay of a K=14 isomer, competing with
stronger branches having small $\Delta$K.
An investigation of the decay modes of isomers
in $^{176}$W at similar spins allows
a comparison of these two isotones within
the framework of the proposed models.  This
paper presents the results of a detailed experimental
study of both the delayed and prompt gamma decay
of $^{176}$W, including the observation of
an isomer with an unusual pattern of decay.
The broad spectroscopic information
contributes vitally to the understanding of
the physics of K-isomers,  by identifying
many of the low-lying intrinsic structures which are
the building-blocks for the multi-quasiparticle K-isomers,
and by mapping out, as completely as possible,
the set of levels available for the decay
of the K-isomers.  Calculations are
also performed and compared with the experimental
data.  Partial results of this work have been
reported previously \cite{Cro93b}.

\section{SPECTROSCOPY OF $^{176}$W}
\subsection{Experiment}

	The experiment was carried out at the ATLAS
facility at Argonne National Laboratory using
the reaction $^{150}$Nd($^{30}$Si, 4n) at
a beam energy of 133 MeV.  The target consisted
of 1.1 mg/cm$^{2}$ of isotopically pure $^{150}$Nd
on a 53 mg/cm$^{2}$ Pb backing, with a thin
layer of Au evaporated on the front of the
target to prevent oxidation.  The intrinsic
timing characteristics of the accelerator
system provided pulses of $<$ 1 ns duration,
separated by 82 ns.  The gamma rays were detected
by the Argonne-Notre Dame BGO gamma-ray facility
\cite{Ye91},  which consists of 12 Compton-suppressed
Ge detectors and an inner array of 50 BGO
elements.  The master trigger for the experiment
required two Compton-suppressed Ge detectors
to fire within 120 ns of each other, in coincidence
with at least four BGO detectors.  In all,
45 $\times$ 10$^{6}$ events
were recorded.

	In order to extract information on weakly
populated K-isomers,  the data from the BGO
detectors were recorded and analyzed in more
detail than is usually required in studies
of high-spin states.  The energy and time
parameters of each individual BGO element
were written to tape on an event-by-event
basis,  in addition to the usual energy and
time parameters of the Ge detectors.  All
times were measured relative to the radio-frequency
signal from the accelerator, which is related
to the time of arrival of the beam pulse at
the target.
The terms ``prompt'' and ``in-beam''
will be used interchangeably to indicate gamma-ray
transitions occurring during a beam burst (as
defined by the timing resolution of the detectors).

	The energy and timing information
from the BGO detectors
was used to create several data sets
of Ge-Ge coincidences with different coincidence
requirements.  In addition to the normal BGO
fold (i.e. the number of BGO elements that
fired within the coincidence window) and BGO
sum energy parameters, a delayed BGO fold parameter
was created, defined as the number of BGO
elements that fired between beam bursts.
This provided an extremely sensitive and efficient
coincidence trigger for isolating decays of
high-spin isomers.  Gamma-gamma coincidence
matrices were
constructed with requirements
on both total and delayed BGO fold in order
to enhance coincidences above,  below,
and across the isomers.

	Spectroscopy of isomers requires an accurate
determination of the timing of the detected
gamma rays.  In addition, an accurate measurement
of the intensity of both prompt and delayed
gamma rays is extremely sensitive to the
requirements on the time parameters.  Therefore,
these parameters need special attention in
order to correct for the intrinsic changes
in timing of the Ge and BGO detectors with
variations in gamma-ray energy.  Constant-fraction
discriminators (CFDs),  which were used for
both Ge and BGO timing,  provide a first-order
correction for the dependence of the timing
signal on the amplitude of the pulse from the
detector.  In
practice,  however,  variations as a function of
energy are observed, both in the mean timing
and the timing resolution, due to the inherent
physics of the detector's operation.
This effect is not automatically
corrected for in an efficiency calibration
using radioactive sources,
for which there is no timing reference and
therefore no time parameter.
Incorrect gamma-ray intensities will
be extracted from coincidence data
if no correction is made.
The timing parameters
for both BGO and Ge detectors were therefore
adjusted off-line according
to the empirical formula
$t \rightarrow t + aE_{\gamma}^{b}$,
where $a$ and $b$ were fit to the data.  When
setting a coincidence condition around the
prompt peak at $t_{0}$ in the Ge time spectra,
it was necessary as well to vary the width
of the coincidence window (and the width of
the window for subtracting random coincidences)
to accept a fixed fraction of the peak, regardless
of energy.  The observed
widths of the peaks in the time spectra,
as a function of energy,  were fitted to the
functional form $w_t = cE_{\gamma}^{d}$,  and
the widths of the time windows were set accordingly
as a function of energy, usually to cover
the range of $t_{0} \pm 1.7 \sigma$,
where $\sigma$ denotes the
standard deviation.

\subsection{Level-Scheme}

	The complete decay-scheme of
$^{176}$W derived
in this experiment is shown in
Fig. \ref{level_scheme}.
It includes 148 transitions,
about a factor of four more than reported
in earlier work \cite{Dra78,Lee84}.  The energies
and intensities of the observed gamma rays
are given in Table I, along with the spin
assignments of the relevant levels.  The intensity
of the weakest gamma ray recorded in the table
is about 0.025\% of the total population of
$^{176}$W, corresponding to an absolute
cross-section of $\sim$5 $\mu$b.
Twelve rotational bands are observed
(Fig. \ref{level_scheme}),
of which eight are new in this
work.  The bands with K$>$0 are labeled in the
level scheme by
numbers from one to nine to facilitate
discussion in the text.  The ground-state
band (states involving only collective rotation)
is referred to as the g band, the lowest configuration
containing two rotation-aligned neutrons
as the s band, and the band built
on the first excited K$^{\pi}$=0$^{+}$ state
as the 0$_{2}{^+}$ band. The spins of the levels
observed in this
work were deduced by the method of Directional
Correlations from Oriented states (DCOs),
with the DCO ratios defined as in Ref. \cite{Ye91}.
These ratios are given below where they are
relevant for determining the spins of the
various band-heads.  Quadrupole transitions were
found to have DCO ratios clustered around
0.95, and dipoles around 0.45.
A more complete list of DCO ratios is
given in Ref. \cite{Cro93a}.
The observation of regular rotational cascades
of M1 and E2 transitions throughout the level-scheme
served as an additional check on the  consistency
of the spin assignments. In the following
two subsections, the level scheme is separately
discussed for (i) the K=0 and intermediate-K states
primarily observed in the prompt spectroscopy,
and (ii) the feeding and decay of high-K isomers.

\subsubsection{K=0 and Intermediate-K States}

	The yrast s-band states in $^{176}$W were
extended to J$^{\pi}$ = 26$^{+}$.  The g band
was also extended beyond the backbend to spin
22, and is discussed in more detail below.
Many weakly populated, non-yrast structures
were first noticed in this experiment because they
were fed preferentially by the decay of high-spin
isomers.  The study of these structures probably
would not have been possible with prompt
spectroscopy alone.
For example, a large number of non-yrast
levels of the g, s and 0$_{2}{^+}$
bands (Fig. \ref{level_scheme}),
were first observed via decay
of the 14$^{+}$ isomer.  These non-yrast structures
were then extended in prompt spectroscopy.
Representative
gamma-ray spectra for the g- and s-band states
are shown in Fig. \ref{k0_spectra}.
The observation
of the non-yrast members of the g band and
s band is noteworthy,  not just because these
states are rarely observed,  but also because
of their relevance for understanding the highly
K-violating decays.  In particular,  detailed
spectroscopy of these bands in the region
where they cross and interact allows the extraction
of a matrix element for the interaction, and,
as discussed in Section IV,  this has important
implications for testing the role of Coriolis
mixing in the decay of high-K states to the
g and s bands.

	The spins of the various K=0
and intermediate-K
rotational bands are all unambiguously determined,
except for band 7,  where the DCO ratio of
0.7(1) for the 1241 keV transition
is not conclusive.
The other two newly observed 2-qp
configurations are the band-heads of
bands 5 and 6, whose spins are determined
from the DCO ratios of 1.3(2) and 0.4(1),
respectively, extracted for the
397 and 1010 keV transitions.
The tentative parity
assignments of bands 5 and 8 are based on the
observation of fast quadrupole transitions
linking states in these bands to
structures with significantly different
values of K.
If these (K-violating) transitions were of
M2 character, unprecedentedly low hindrance
factors would be implied (e.g.,  F$\leq$1
for the 397 keV transition depopulating band
5,  compared to 10$^{6}$-10$^{10}$ for a typical M2
transition with $\nu$=6 \cite{Lob68}).  The
negative parity tentatively assigned to band
3 (and consequently, to its signature partner,
band 4) is based on systematics.  This parity
assignment is also natural in view of the
pattern of decay of band 3 to the g band,
in which no quadrupole transitions are observed.
For the 397-441
keV sequence near the bottom of band 5,  the
proposed ordering of transitions is based
on the rotational pattern of levels.

\subsubsection{Feeding and Decay
of High-K  Isomers}

	Two new high-spin isomers were observed in
$^{176}$W (Fig. \ref{level_scheme}).
The more strongly populated one,
for which complete patterns of feeding and
decay were established, is
assigned K$^{\pi}$=14$^{+}$
for reasons explained below.
Delayed feeding of yrast
states, pointing to the existence of such
an isomeric level in $^{176}$W, had previously
been reported \cite{Dra78}.  An indication
of the quality of the data can be gained from
the spectra in Fig. \ref{dd_spectra}.
The decay of
the isomer, with a measured half-life of t$_{1/2}$
=35 $\pm$ 10 ns, includes a strongly K-violating
branch,  connecting the isomer directly to
the 14$^{+}$ member of the ground-state band,
as well as other direct branches to levels
with K=0.  In a previous description of this
work \cite{Cro93b}, the half-life of this
isomer was stated as $\sim$70 ns.
An accurate measurement
of the half-life was difficult in this
experiment, since it was not possible
to use the delayed BGO fold parameter to enhance
the peaks whose half-lives were being fitted,
since this would have introduced a bias
on the time of the decay.  In a more recent experiment
using the reaction $^{50}$Ti($^{130}$Te,4n)
and a recoil-shadow geometry \cite{Cro95},
the Ge detectors were shielded from
gamma rays emitted directly from the target,
but were able to detect transitions emitted
from recoiling $^{176}$W nuclei that stopped
in a Pb catcher foil 17 cm downstream.
An event was collected if two Ge detectors
and two BGO detectors registered signals
in coincidence.
Consequently, the peak-to-background ratios
of the transitions depopulating the isomer were
enhanced by more than an order of magnitude
without any use of the delayed BGO fold parameter,
thus allowing a much more reliable determination
of the half-life, as shown in Fig. \ref{half_life}.
The most significant cause of deviations from the
fits lies in the systematic error in choosing
the correct amount of background-subtraction for
the relevant gamma-ray peak.  Other effects, such
as the presence of later beam pulses and the feeding
of the 35-ns isomer by a higher-lying isomer, are
less significant.  Despite these uncertainties, the
half-lives extracted from the 714, 491 and 408 keV
transitions, as shown in Fig. \ref{half_life},
as well as the 558, 440 and 917 keV transitions,
all cluster around a value of 35 ns,
within the present error bars of  $\pm$10 ns.
Table II gives the intensities
of the transitions depopulating the 35-ns
isomer.

	The delayed BGO fold parameter proved to
be a very powerful tool for investigating
the decay of the 35-ns isomer.  Although the isomer
was populated in this reaction with an
intensity of only 2\% of the $^{176}$W
reaction channel,
the gamma rays associated with the decay
constituted 15\%
of all events with delayed BGO fold $\geq$ 2,
and 45\% of those with delayed fold $\geq$
3.  The power of this method is illustrated
by a comparison (Fig. \ref{kdel_spectra})
of total projections
from the coincidence matrices,  both with
and without a coincidence requirement that
the delayed BGO fold be at least four.  The isomeric
decays are not visible above the background
without the delayed fold condition, but are clearly
visible after the condition is imposed.

	In addition to the 35-ns isomer,
a second isomer with t$_{1/2}$ $\approx$
10 ns was observed (Fig. \ref{level_scheme}).
The DCO ratios of the transitions
depopulating the 35-ns and 10-ns isomers
were not measurable,  due to the very low
intensities involved,  or deorientation of
the nuclear spin during the lifetimes of the
isomers,  or both.  The interlocking pattern
of branches in the decay of the 35-ns isomer
restricts its spin to J$\leq$14,  assuming
that no $\lambda$=3 transitions are involved.  For
example, if the 917 keV transition were of
E3 character,  its strength would be 17 W.u.,  while
E3 transitions with
a degree of K-forbiddenness, $\nu$,
as low as 1-2 typically
have strengths of 10$^{-2}$--10$^{-5}$ W.u.
\cite{Lob68}.  Direct decays are observed
from the 35-ns isomer to the 14$^{+}$ states
of the g and s bands, but not to the
12$^{+}$ levels.  This suggests J$>$13
for the spin of the isomer,  since for J$\leq$13
one would expect stronger branches to the
12$^{+}$ states,  based on the associated
multipolarities and gamma-ray energies.  The
M1/E2 branching ratios for the transitions
in the band built on the isomer
and the absence of any rotational alignment are
also consistent with those expected for the
lowest K=14 deformation-aligned configuration,
as discussed in Section III.  Thus, the 35-ns
isomer is assigned J$^{\pi}$ = 14$^{+}$.

	Decay schemes were also constructed for the
rotational bands built on the two isomeric
band-heads.  The technique used to enhance
these cascades was the requirement of a
delayed coincidence with gamma rays from the
decay of the 35-ns isomer,  before inspecting
the data for coincident transitions with prompt
timing.  These bands were then investigated
further using prompt-prompt coincidence matrices
constructed with the requirement of a high
delayed BGO fold.

	The ordering of the following cascades in
the decay of the 35-ns isomer
(Fig. \ref{level_scheme})
could not be determined:  884-656,  445-1096,
and 519-1068-512 keV.  The
ordering of the 316-210-267-355 keV cascade
depopulating the 10-ns isomer is based on
the intensities of the transitions;  the time
spectra gated on these transitions are also
consistent with this ordering.

	A band was also observed built on a high-K
state tentatively labeled K=(13).  This band
(band 9 in Fig. \ref{level_scheme}) is very prominent
in spectra gated on very high delayed fold
and is thus apparently populated partly by the
decay of another isomer with even higher spin
(Fig. \ref{level_scheme}).  The decay pattern of the
K=(13) band-head could not be mapped out entirely
because of its fragmentation into many weak
branches,  nor could the decay scheme of the
very high-spin isomer feeding it be constructed.
 In contrast to the 14$^{+}$ isomer,  the
decay of the K=(13) state appears to have
a half-life too short to measure with electronic
timing,  and seems to occur both to K=0 states
and to the bands with intermediate K values,
such as band 8.

\section{DISCUSSION}

	The most important result of this experiment
is the observation of highly K-violating transitions
in $^{176}$W leading directly from the K=14 isomer
to the K=0 states in the g,  s
and 0$_{2}{^+}$ bands.  As noted in section I,
there have been several recent observations of transitions
with anomalously large values of  $\Delta$K in neighboring
nuclei in this region.  In all these previous
cases \cite{Cho88,Hf174,Sle90,Kra88,Kra89},
however, the anomalous decays constituted only a
small fraction of the total decay of
the isomer, while the predominant path of
deexcitation was one that followed the usual K-selection
rules.  An exception is the decay of an isomer
in $^{179}$W \cite{Wal94a}, which involves
an accidental degeneracy \cite{Wal94b,Cro94}.
In contrast, the novel feature in $^{176}$W
is that the {\em majority} of the
decay of the K-isomer proceeds
through highly K-violating $\Delta$K=14
transitions, with no observable decay to the
many available states with intermediate values
of K.

	The identification of a
K=14 isomer in $^{176}$W
with multiple K-violating decay branches is
especially interesting because similar K-violating
decays have recently been studied in the neighboring
even-even isotone $^{174}$Hf \cite{Hf174}.
In addition,
as discussed below, we propose a quasiparticle
configuration for the isomer in $^{176}$W
that is the same as that of the isomer in
$^{174}$Hf.  The ability to compare hindrance
factors for the decay of K-isomers with the
same configuration in different nuclei
makes this pair of isomers a powerful tool
for exploring the mechanisms of highly K-violating
decays,  since it is possible to examine the
systematics of the decays while keeping many
of the relevant degrees of freedom constant.
The spin and K assignments for the states
to which the isomer decays, as well as the
levels to which
no decay branches are observed,
are essential to the discussion of variation
of the hindrance factors with K-forbiddenness.
Therefore, the K=0 and low-K
states are discussed first, followed
by an analysis of
the feeding and decay of the high-K isomers.

\subsection{Rotational Bands With K=0}

	Fig. \ref{ex} shows the relative excitation
energies of the observed K=0 states as a function
of J(J+1).  From the figure, the presence
of three rotational bands with two band-crossings
is obvious.  The non-yrast parts of the g
and s bands are easily identified from this
figure, and these assignments are confirmed
by the deviations from the expected rotational
energies in the crossing region, as well as
the branching ratios between the two
bands.  In a simple two-band-mixing scenario,
assuming that the interaction
matrix element, V$_{gs}$,
between the two bands is constant with
respect to spin, an upper
limit of $\mid$V$_{gs}$$\mid$$<$33
keV can be stated from the 66-keV difference
between the 16$_{g}^{+}$ and 16$_{s}^{+}$ levels,  since the
energy separation is always greater than 2V$_{gs}$.
The best source of a lower limit on V$_{gs}$
comes from the interband branching ratios, rather than
from the energy levels.  Such branching ratios
have seldom \cite{Kho73a} been measurable
in previous experiments.  In Table III, the
measured ratios, B(E2)$_{out}$/B(E2)$_{in}$,
for the $\Delta$J=2 transitions within and
between bands are
compared to simple two-band-mixing calculations
\cite{Kho73a} for different
values of $\mid$V$_{gs}$$\mid$
, where ``out'' refers to the out-of-band
branch (g$\rightarrow$s
or s$\rightarrow$g) and ``in'' to
the in-band branch (g$\rightarrow$g
or s$\rightarrow$s).
The best agreement between calculations and
experiment is obtained
with $\mid$V$_{gs}\mid \approx$ 30 -- 33 keV.
The ratios
are largest near the crossing of the two bands
(between J=16 and 18),  where the mixing is
the greatest. Although the measured values
do not all overlap with the calculated numbers
within statistical errors, it should be noted
that while the ratios span two orders of magnitude,
the values calculated under the simple model
assumptions mentioned above are within a factor
of 5 of the experimental numbers.
A value of $\mid$V$_{gs}$$\mid$$\approx$32
keV is adopted below for the discussion of
possible mechanisms for the K-violating decays.

	At high spins, a large but gradual increase
is observed in the moment of inertia of the
g band, as evidenced by the curvature
of its E versus J(J+1)
trajectory in Fig. \ref{ex}.  Similar effects
were observed,  e.g., in $^{179}$W \cite{Wal91},
 where the yrast band was also observed to
high rotational frequencies.  When this type
of gradual increase is observed,  it has been
shown \cite{Ben86b,Ben89c,Ben91}
that it is not possible,  based only on the observed
energy levels,  to separate
the contributions
of centrifugal stretching,
Coriolis anti-pairing,  and gradual quasiparticle
alignments.  Calculations using the cranked
Nilsson-Strutinsky code of Ref. \cite{Ben89e}
have been carried out as part of the present
work.  These calculations
indicate that the deformation of
the g band increases gradually from
($\epsilon$$_2$)$_{g}$=0.24
in the ground-state to
($\epsilon$$_2$)$_{g}$=0.29
at spin 18, and then remains approximately
constant from spin 18 through spins greater
than 30.  (The
details of the calculations are given in
subsection C.)  This
increase in deformation will lead to an increase
in the moment of inertia,  with some contribution
as well from the Coriolis anti-pairing effect.
It should be noted that no alignment process
other than the g-band/s-band crossing is predicted
to occur in the range of frequencies being
considered.  (Alignment processes in this
region have been discussed elsewhere
\cite{Ben91,Ber76}.)  No strong centrifugal stretching
effect is predicted for the s band,  which
is expected to maintain a deformation of
($\epsilon$$_2$)$_{s}$$\approx$0.22--0.23
for the range of spins observed in this experiment.
This is in agreement with the data,  since
the s band does not show any deviation from
J(J+1) level-spacing (Fig. \ref{ex}),  such
as would be expected if there was significant
stretching.  Near the g-band/s-band crossing,
a rather large difference in deformation
is predicted between the s band
and g band [($\epsilon$$_2$)$_{s}$=0.22,
($\epsilon$$_2$)$_{g}$=0.29], consistent
with the observation of only a small interaction
between the bands.  Such deformation changes
may be partially responsible for the complicated
alignment curves observed in the yrast sequences
of the W isotopes \cite{Ben91,Ber76}.
It has also been suggested that quadrupole
pairing plays a role \cite{Sun94} in the alignment
processes in this region, although the effect
is predicted to be far more pronounced in
the odd-Z systems.

	It has been debated
\cite{Wal94b,Cro94} whether
the structure referred to here as the s band
might in fact involve two quasineutrons
coupled in the novel Fermi-aligned
scheme \cite{Fra93a}, in which the quasineutrons
create a state with non-zero
$<$K$>$ = $<$J$_z$$>$, as
well as a non-zero rotational alignment,
$<$J$_x$$>$.  This coupling scheme
is predicted to arise from a balance
between the mean field and the Coriolis
force, and is stable for certain ranges
of values of
the rotational frequency, pair gap, and
Fermi level.  This coupling scheme, however,
cannot apply to the band presently being
discussed for the following reason.
The band shows a nearly perfect pattern of
J(J+1) level-spacings (Fig. \ref{ex})
from spin 12$\hbar$ to spin 26$\hbar$,
implying a constant
rotational alignment.
Thus, only one coupling scheme
is involved in this range
of spins.  Since,
even in the most favorable
cases, the Fermi-aligned coupling
scheme is not predicted \cite{Fra93a} to
be stable at rotational frequencies
as high as those
observed here ($\omega$ $>$ 0.35 MeV),
the entire
sequence of states is assigned to
a rotation-aligned configuration
(``s band''), with
$<$K$>$ = 0.

	A third band populated by the decay of the
isomer is assigned K=0.  It is identified
as the band structure built on the first excited
0$^{+}$ state, and labeled as the 0$_{2}{^+}$
band, based on the following observations:
(a) In the region where the 0$_{2}{^+}$
band crosses the s band,  the measured perturbations
of the 12$^{+}$ level energies and the inter-band
branching ratios give
$ \mid V_{0_{2}^{+},s} \mid$
$\sim$ 30--110 keV for the interaction matrix
element.  (Sufficient data for a more
detailed calculation
such as the one performed above
for $\mid$V$_{gs}$$\mid$
are not available.)  An interaction this large
would be quite anomalous for bands with different
K values.  (b)  No signature partner band
is observed in the data.  (c) The electromagnetic
transitions to the g band are fairly strong
(B(M1)$\sim$0.02 W.u.,  assuming a mixing
ratio of $\delta$=0) compared to those normally seen
\cite{Lob68} for K-violating transitions.
(d)  Although the depopulation of the band
at low spins is too sudden to allow the observation
of levels with J$<$6 in these data,  two states
in $^{176}$W have been observed (at 929 and
1117 keV) in the radioactive decay of the
3$^{+}$ state in $^{176}$Re \cite{Ber77}, which form
a clear continuation of the J(J+1) pattern
of excitation energies (see Fig. \ref{ex}, where
these levels have been included).  Furthermore,
the levels from 2$^{+}$ to 10$^{+}$ of the
0$_{2}{^+}$ band in $^{176}$W are nearly
identical (gamma-ray energies differing by
$<$8 keV) to the corresponding levels of the
0$_{2}{^+}$ band in $^{178}$W \cite{Soo91},
which has been observed down to the K=0 bandhead.
 The log ft values are very similar (log ft=6.7
and 6.9,  respectively \cite{Ber77,Gou70})
for population of
the corresponding 2$^{+}$ states in $^{176}$W
and $^{178}$W from the $^{176}$Re
and $^{178}$Re $\beta$-decays.

	The observation of the crossing between the
s band and the 0$_{2}{^+}$ band is noteworthy,
since it has only been observed in a few
cases \cite{Kho73a,And74,Lie74}.
The band built on the first excited 0$^{+}$
state is often labeled as the $\beta$-vibrational
or ``$\beta$-band.''  The extrapolated band-head energy
of 0.9 MeV is considerably lower than that
expected for a 2-quasiparticle (qp) state,  but attempts
to associate such states in this region with
simple collective modes like $\beta$-vibrations
or pairing vibrations have failed to reproduce
the available data on electromagnetic and
particle-transfer matrix elements
\cite{Mik67,Bes66,Bir75,Bur88}.
We therefore avoid the label ``$\beta$-band'' in this
paper.  The measurement of the interaction
V$_{0_{2}^{+},s}$ in this work may test future
theoretical descriptions of these low-lying
excitations.

\subsection{States With Intermediate Values of K}

	A large number of intermediate-K states is
observed in this experiment
(bands 3-8 in Fig. \ref{level_scheme}),  and
most of these can be assigned 2-qp configurations
with some confidence.  In this section, we
discuss the structures of these states, and
compare their properties with the results
of cranked Woods-Saxon calculations performed
using the computer code and the ``universal''
parameters described in Ref. \cite{Cwi87}.
Cranking calculations can only determine
angular momenta and Routhians relative to
some additive reference.
We have chosen as the reference a
function $\Im = \Im_{0} + \Im_{1}\omega^{2}$,
with $\Im_{0}$=43 $\hbar^{2}$ MeV$^{-1}$ and
$\Im_{1}$=80 $\hbar^{4}$ MeV$^{-3}$
fitted to the moment
of inertia of band 6 as a function of
rotational frequency,
because this band is observed over
a large range of rotational frequencies.

	Bands 3 and 4 require a short preface since
it is not entirely clear that 2-qp assignments
are appropriate for these bands.  Based
on previous data,  it had been suggested \cite{Dra78}
that these sequences might correspond to two
of the four members of the octupole-vibrational
multiplet with
K$^{\pi}$=0$^{-}$,1$^{-}$,2$^{-}$,3$^{-}$. Such
an interpretation would require the existence
of band-heads with K$<$4, which might not have
been observed in previous experiments purely
because of the early depopulation of these
bands into the g band.  The present, more sensitive
experiment has,  however, failed to reveal
any further in-band transitions leading to
bandheads with K$<$4,  calling
the vibrational interpretation  into question.
As shown below,
many of the characteristics of these bands
can be understood naturally if they are the
two signature partners of a two-quasiproton
band with K=4.  We will briefly note below some of
the successes and failures of such a description.

	A first step in classifying
the intermediate-K
band-heads as two-quasiproton ($\pi$$^{2}$)
or  two-quasineutron ($\nu$$^{2}$) states
is to examine the systematics of the band-head
energies in nearby nuclei
(Fig. \ref{syst_2qp}).
The K= 4$^{-}$ and
7$^{-}$ states occur at approximately constant
excitation energies in the W isotopes with
A=172--184,  suggesting that they are
$\pi$$^{2}$ states.  The energies of the 6$^{+}$
states in the Hf isotopes show evidence of
a minimum near N=104;  this is in agreement
with calculations of the quasi-neutron states
(see below),  which show that the configuration
$\nu$ 5/2$^{-}$[512]  $\otimes$
$\nu$ 7/2$^{-}$[514] should
be favored in this vicinity,  rather than
$\pi$$^{2}$ configurations.  (Less extensive
data are available for the 6$^{+}$ states
in the W isotopes.)  Both $\nu$$^{2}$ and
$\pi$$^{2}$ states with K$^{\pi}$ = 8$^{-}$ have
been observed in the Hf-W region,  but for
$^{176}$W the $\pi$$^{2}$ state is predicted
to be much lower in energy
than the $\nu$$^{2}$ state
based on the present calculations.  Given
these inferences regarding the $\pi$$^{2}$
and $\nu$$^{2}$ characters of the states,
the configurations are unambiguously determined
by the available combinations of $\Omega$ values,
as supported both by the present Woods-Saxon
calculations and the observed 1-quasiparticle
states in the neighboring odd-A nuclei.
The predicted excitation energies are
compared with experiment in Table IV.  The
agreement of the calculated and measured excitation
energies is satisfactory,  except in the case
of the 4$^{(-)}$ state.

	Fig. \ref{routhians} compares
the experimental 2-qp
Routhians with the calculated ones.  The slopes
of the observed Routhians are well reproduced,
as are the relative energies of the 6$^{+}$,
7$^{-}$,  and 8$^{-}$ Routhians.
While the two 4$^{-}$ Routhians
are calculated to be considerably higher
in energy compared to experiment, they show
the correct energies relative to one another
as well as the correct slopes and alignment
behavior.  The calculations predict
the excitation energy of the 6$^{+}$ intrinsic
state to be lower and the
signature splitting of the 4$^{-}$ Routhians
to be larger than the experimental observations.

	No attempt is made here to discuss band
7 in detail,  since its parity and
K are unknown.

\subsection{High-K States}

	In the following discussion, the
data for the high-K states are compared
with the results of cranking calculations.
The Woods-Saxon potential, as described above
for the intermediate-K, 2-qp states,
is used wherever possible because of its somewhat
better description of the structure of the
quasiparticle levels.  The modified harmonic
oscillator (Nilsson) model was used, however,
for the calculation of potential energy surfaces
in Section IV, since a computer code was available
\cite{Ben89e} which could transform the calculation
from the rotating frame into the laboratory
frame.  This allowed potential energy surfaces
to be constructed for a fixed value of $<$J$_{x}$$>$
rather than for a fixed rotational frequency.
 For these calculations, described in more
detail below, the Nilsson parameters of reference
\cite{Ben85} were used.

	To understand the unusual pattern of decay
of the 14$^{+}$ isomer, it is important to
analyze both the decay of the isomer and the
rotational band built on it in order to obtain
as many clues as possible regarding its underlying
structure.  Both the Woods-Saxon and Nilsson
calculations show that based on the $\Omega$ values
of the states lying near the Fermi surface,
a state with a value of K this large cannot
be constructed from the angular momenta of
only two quasiparticles.  The 14$^{+}$ isomer
is therefore
most likely to be a 4-qp state,  since 6-qp
states are not predicted to occur this low
in energy.  Band 1, the band built on the
isomer, is composed of two signature partners
with no signature-splitting, implying that
no $\Omega$=1/2 orbitals are occupied.
Band 1 is observed up to the
rotational frequencies where
the i$_{13/2}$ neutron rotational alignment occurs
in the yrast cascade,  and neither signature
partner shows any sign of rotational alignment.
This strongly suggests that the i$_{13/2}$ alignment
is blocked by the intrinsic configuration.
The 4-qp configuration predicted
by Woods-Saxon calculations to lie closest
to the yrast line is

K$^{\pi}$=14$^{+}$,
$\pi$ 7/2$^{+}$[404] $\otimes$
$\pi$ 9/2$^{-}$[514] $\otimes$
$\nu$ 7/2$^{+}$[633]
$\otimes$  $\nu$ 5/2$^{-}$[512]

The observed
B(M1)/B(E2) branching ratios agree very well
(Fig. \ref{m1_e2}) with those calculated  according
to the strong-coupling formulae of Ref. \cite{Don83},
using quasiparticle states with the above
Nilsson quantum numbers and spin g-factors
equal to 60\% of their free-space values.
This configuration has a neutron
i$_{13/2}$ orbital occupied,  consistent with the
absence of rotational alignment.
The calculated excitation energy is 3.5 MeV,
in good agreement with the observed value
of 3.746 MeV.

The spectroscopic information for band 9
(Fig. \ref{level_scheme}) is rather incomplete.
A comparison of its gamma-ray energies with those
of the band built on the 35-ns isomer (band 1),
suggests a spin of roughly 13.
The Woods-Saxon calculations predict a near-yrast
state with the configuration K$^{\pi}$=13$^{+}$,
$\pi$ 5/2$^{+}$[402] $\otimes$ $\pi$ 9/2$^{-}$[514] $\otimes$
$\nu$ 7/2$^{+}$[633] $\otimes$ $\nu$ 5/2$^{-}$[512].
The B(M1)/B(E2) data for this band are not very
precise,  but seem to be consistent with those
calculated for this configuration.  (The data
are not consistent with switching the configurations
of the 13$^{+}$ and 14$^{+}$ bands,  since the
calculated B(M1)/B(E2) values differ by an
order of magnitude.)

Little can be said regarding the configuration
of band 2, since the spin and
parity of the 10-ns isomer, on which the
band is built, are unknown.

\subsection{Summary of Data on the Decay of the High-K
States}

	The most striking observation in this experiment
is that the decay of the 14$^{+}$ isomer deviates
completely from the normal pattern of decay
of a K-isomer.  The majority (52\%) of the
decay proceeds directly to K=0 states,  and
no branches of any detectable strength were
found to the states with intermediate K values,
 even though many such levels are available,
as shown above.  This is all the more remarkable
since a K=14 isomer  with the same quasiparticle
assignment has also been observed in the neighboring
even-even isotone $^{174}$Hf
\cite{Hf174,Sle90},
but in $^{174}$Hf
the highly K-violating branch  represents
only a small fraction ($\sim$2\%) of the total
decay.

	Since detailed data are available on the
half-lives and branching ratios for the anomalous
decays of the K=14 states with identical
quasiparticle
assignments in $^{176}$W and $^{174}$Hf, both
absolute and
reduced hindrance factors (f=F$^{1/\nu}$) for the
$\Delta$K=14 transitions in these two
nuclei are compared in Table V.  (Transitions
have been included in the table only if they
are significant at the two-sigma level.)
Although the differences between the reduced
hindrance factors of the two nuclei may
appear small,  they represent very large changes
in the absolute hindrance factors.  For the
decays of the 14$^{+}$ states to the g bands
in $^{176}$W and $^{174}$Hf,  for example,
the absolute hindrance factors differ
by more than two orders of magnitude.
In the following section,
the focus is on trying to understand this
large difference in absolute hindrance factors.

\section{CALCULATIONS FOR
DECAYS OF HIGH-K ISOMERS}

	Since the first observations
\cite{Cho88,Hf174,Wal94a}
of gamma-ray transitions between states with
large values of $\Delta$K but
anomalously low hindrances compared to the
usual K-selection rules, it has become necessary
to reevaluate old assumptions about K-violation,
as one would like to understand the mechanism
responsible for these exotic decays.  Since
the decays arise from small-probability fluctuations
in the K quantum number, it is not possible
to gain direct insight into the decay mechanisms
by measuring quantities
such as excitation energies or electromagnetic moments
and transition rates.  At present, the mechanisms
of K-violation can only be addressed by comparing
the K-violating transition strengths
with theoretical calculations.

	In this section, the focus
is on understanding
why differences of several orders of magnitude
exist between the strengths of the $\Delta$K=14
transitions in $^{176}$W and $^{174}$Hf.
As discussed in section I,
two suggested mechanisms that might be responsible
for coupling states with different values of K
are $\gamma$-tunneling \cite{Cho88,Ben89d}
and Coriolis mixing \cite{Hf174}.  The
Coriolis force mixes wave-functions
corresponding to the same shape but oriented
differently with respect to the angular momentum
vector.  In contrast, $\gamma$-tunneling
involves large-amplitude collective motion
in which the shape of the nucleus changes.

	The distinction between these two modes of
K-mixing can be tested empirically:
Coriolis mixing should manifest itself in
systematic variations in K-mixing as a function
of configuration and rotational frequency,
while $\gamma$-tunneling models predict that
K-violating decays should show strong variations
in strength as a function of the height and
shape of the barrier in the potential energy
surface. Both concepts are discussed below in light
of the present data.

\subsection{Testing predictions of Coriolis mixing}

	As described above,
an explanation previously
proposed to account
for the K-violating decays
in $^{174}$Hf \cite{Hf174}
associates a change in Coriolis mixing at
high spins with a change in the structure
of the yrast states from g-band to s-band
configurations.  The mean K-value of the s-band
levels is zero, but because the two quasi-neutrons
are aligned perpendicular to the symmetry
axis, admixtures of various K values are to
be expected in the s-band wave-function, with
values of K ranging from 0 to approximately
$\pm$K$_{max}$,
where
K$_{max}^{2} = j_{max}^{2} - i^{2}$.
Here,
$j_{max}$ is the maximum spin to which the
two neutrons can be coupled subject to the
Pauli exclusion principle, and $i$ is their
rotational alignment.  The alignments of
the s bands in $^{176}$W and $^{174}$Hf
extracted from the data are
8.7$\hbar$ and 6.0$\hbar$, respectively,
leading to K$_{max}$
= 8.3 and 10.4.  (We use $j_{max}^{2}$, rather
than $j_{max}(j_{max}+1)$, since this is
only an approximate treatment.)
The previously observed highly K-violating
decays are then ascribed to Coriolis mixing,
with essentially normal values of the hindrance
per degree of K-forbiddenness, i.e. f$\sim$100,
but with a smaller degree of K-violation,
$\nu^{\prime} = \nu - $K$_{max}$.  Although
decays are also
observed to the g-band states,  such decays
are attributed in this model \cite{Hf174}
to mixing of
the g- and s-band configurations in the band-crossing
region, so that the hindrance factors should
be approximately
$F=f^{\nu^{\prime}}/(\mid<\Psi\mid s>\mid^{2}
\times \mid<s \mid K_{max}>\mid ^{2})$.
Here,
$\mid <s \mid K_{max}>\mid^{2}$
is the squared amplitude
of the part of the pure s-band wave-function
having K=K$_{max}$, and
$\mid <\Psi \mid s>\mid^{2}$
is the squared
amplitude of the s-band configuration in the
wave-function of the state in question, as
calculated in Section III.

	The model predicts a
strong correlation between the hindrance factor
and the amount of s-band admixture in the
final state. The smaller value of K$_{max}$
in $^{176}$W should also contribute to increased
hindrance factors relative to those in $^{174}$Hf.
Fig. \ref{coriolis} presents a test of the predictions
of this model for the two 14$^{+}$ isomers
in $^{174}$Hf and $^{176}$W.  Although the
hypothesis was a reasonable one to account
for the $^{174}$Hf data,  the addition of
the $^{176}$W data to the figure reveals that
there is essentially no correlation between
the s-band admixture and the hindrance factors.
Furthermore, the hindrance factors in $^{176}$W
are orders of magnitude too low, even with
the most drastic assumption of 100\% admixture
of K=K$_{max}$ in the s-band states.  This
is undoubtedly a very schematic treatment
of Coriolis mixing;  the objective was not
to calculate the effect, but simply
to search for the expected trends in the data
and to evaluate whether the orders of magnitude
were in the right range.  The lack of correlation
between the hindrance factors and the s-band
admixtures, however, seems very difficult
to explain in any model of Coriolis mixing.

\subsection{Calculations of $\gamma$-Tunneling}

	The failure of the Coriolis mixing picture
leads one to consider a possible interpretation
of these decays in terms of a $\gamma$-tunneling
picture.  The mechanism is schematically illustrated
in Fig. \ref{tunneling_schematic}.
Dynamic motion in the
shape degrees of freedom of nuclei is often
very difficult to describe theoretically,
as small-amplitude shape vibrations are usually
coupled strongly to the single-particle degrees
of freedom.  Tunneling processes, therefore,
could be extremely important, because they
represent a possible simplification of the
dynamics.  When a nucleus tunnels through
a classically forbidden region along a path
parametrized by a deformation $\gamma$,
the wave-function
is attenuated exponentially, falling off by
1/e over a distance
$\delta\gamma$ $\propto$ $(V-E_{0})^{-1/2}$,
where $E_{0}$ is the energy eigenvalue of
the state, and the potential, V($\gamma$), is defined
as the minimum adiabatic value
of the energy at any value of $\gamma$.
To give some perspective, it is worth
nothing that even when
the decay of a K-isomer occurs with an abnormally
low hindrance, the hindrance factor F is still on
the order of 10$^{4}$--10$^{6}$ or
larger \cite{Cho88,Hf174,Wal94a}.
Thus, if tunneling is responsible for these
decays, the attenuation is extremely severe.
Any motion through non-optimal single-particle
configurations with E$>$V should, therefore,
be strongly attenuated, with the attenuation
length reduced to
$\delta\gamma$ $\propto$ (E-E$_{0}$)$^{-1/2}$.
Tunneling processes may be the only type
of collective nuclear motion other than rotation
for which an adiabatic approximation is justified.
For an adiabatic quantum-mechanical tunneling
process, the only quantities needed to calculate
the tunneling probability are the potential,
V, and the inertial parameter, D, which corresponds
to the mass in the Schr\"{o}dinger equation
and is discussed in more detail below.

	In the present work,
the potential energy has been calculated
as a function of $\gamma$ for
the K=14 isomers in $^{174}$Hf and $^{176}$W
with the cranked Nilsson-Strutinsky method,
using the computer code described in Ref.
\cite{Ben89e}.  In these calculations,  the
deformation parameters $\epsilon$$_2$ and
$\epsilon$$_4$,  as well as the static pair
gaps $\Delta$$_{p}$ and $\Delta$$_{n}$,  were
all varied self-consistently,  i.e. the
$\gamma$-tunneling
path is assumed to be the path of steepest
ascent to the saddle-point and steepest descent
from the saddle-point.  The cranking frequency
was varied to provide a constant value of
$<$J$_{x}$$>$,  and particle-number projection
was employed.  A step-size of 10$^{\circ}$ was used
for $\gamma$.  Considerable variation was found in
$\epsilon$$_4$ as a function of $\gamma$,
with values of $\epsilon$$_4$
ranging from 0.00 to 0.05,  and it
was found that a proper self-consistent
variation of $\epsilon$$_4$ was necessary
for an accurate determination of the potential
energy surfaces.
The parameters $\epsilon$$_2$,  $\Delta$$_{n}$
and $\Delta$$_{p}$ showed variations as functions
of $\gamma$, but remained within about 10\% of their
average values.   The calculated potential
energy curves are shown in Fig.
\ref{potential_energy}.
A summary of some of the calculated parameters
is given in Table VI.

	In an adiabatic model, the information on
the dynamics is contained in the inertial
parameter.  At present, the dynamics are not
understood well enough to allow an accurate
calculation of this quantity from first principles.
One fact that seems to emerge from all microscopic
approaches \cite{Ben89d,Bar90}  is
that superfluidity in nuclei has an important
effect on the inertial parameters.  Another
approach is to take a phenomenological functional
form for the inertial parameter and to fit
adjustable parameters to the available information
on properties of nuclei.  The results of such
an approach \cite{Mol81} fall within the
range of values calculated from the microscopic
methods.  It is worth noting,  however,  that
the starting point for such a fit is the
observed transition rates for tunneling processes
as observed in the decays of
fission isomers \cite{Mol81}.
{}From this
point of view, experiments involving tunneling
processes could directly measure the inertial
parameters.  Such a program might be a reasonable
one to pursue for K-isomers in the long run,
but first it must be established whether a
$\gamma$-tunneling approach can provide a reasonable
description of the data.  This is the more
modest goal of this paper.  As is explained
in more detail below, the calculations factor
out most of the effect of the inertial parameter,
and the final results are inspected to see
whether they can approximately reproduce the
data for a reasonable choice of the inertial
parameter.

	The WKB approximation for the tunneling
probability is
$T=\exp[-2\hbar^{-1} \int \sqrt{2D(V-E_{zp})} d\gamma ]$
where $E_{zp}$ is the zero-point energy,
i.e. the eigenvalue of the Schr\"{o}dinger
equation, and D and V are defined as above.
The tunneling probability, T, represents
the squared amplitude of the part of the wave-function
localized in the potential well
around $\gamma$=0$^{\circ}$,
which is interpreted as the admixture of
the K=0, J=14 component in the wave-function
of the J=14 state (which has predominantly
K=14).
The hindrance factor for the K-violating
transition is calculated as $F^{calc}
= 1/TS$,  which can be compared with the measured
quantity  $F^{exp} =  t_{1/2}/t^{W}_{1/2}$,
as defined in section I.
Here, the absolute normalization of the
transition rate, via the constant $S$,
is determined by the requirement
that $T$=1 should produce values of
$F=1/S$ that are typical
for unhindered transitions, i.e. those
with $\Delta$K less
than $\lambda$, the multipolarity of the phonon.
The strengths of unhindered M1 and E2
transitions typically cover a range of
0.03--0.3 W.u. and 10-100 W.u., respectively
\cite{Lob68}, and the corresponding range of
S is depicted as a horizontal bar in the
figures below showing the final results
of the calculations.
An estimate of the
zero-point energy, $E_{zp}$, was obtained
by calculating the energy for Gaussian trial
wave-functions $\Psi(\gamma)$,  with the Gaussian width
used as a variational parameter.  Since the
isomers are $\sim$1 MeV higher in energy than
the yrast levels with the same spin,  it is
assumed that the dominant contribution to
the decay comes from tunneling from the minimum
at $\gamma$=--120$^{\circ}$ to the one
at $\gamma$=0$^{\circ}$ (giving low-K
admixtures in the initial state),  rather
than from tunneling in the opposite direction
(which would result in high-K admixtures in
the final state).

	As explained above, the philosophy adopted
here is one of extracting the inertial parameter,
$D$, and comparing it with various a priori
estimates.  There are two ways in which $D$,
which is assumed to remain constant with $\gamma$,
affects the calculated hindrance factors.
The more important one is the explicit dependence
of the tunneling probability on $D$.  In addition,
the zero-point energy also depends on
$D$ ($E_{zp} \sim \sqrt{D}$
for a harmonic well).  Neglecting the latter
for the moment, simple algebra yields
$\ln F^{calc} = (\sqrt{D} \times A) - \ln S$,
where the quantity
$A=2\hbar^{-1} \int \sqrt{2(V-E_{zp})} d\gamma$
is independent of $D$.  In other words,
in a log-log diagram of $F^{exp}$
vs $F^{calc}$, the slope is proportional
to $D^{1/2}$,  and a change in $S$
shifts the data-points
horizontally.  The slope equals 1 for the
correct value of $D$, and this is essentially
the method used here to extract values of
$D$ from the data, although the minor dependence
of $E_{zp}$ on $D$ has also been taken into
account, as shown below.

	To describe consistently the variations of
the tunneling probabilities from nucleus to
nucleus,  it is necessary to recognize that
a systematic variation in the inertial parameter
is to be expected, depending on actual deformations
and pair gaps.  Various microscopic models predict
a variation of the inertial parameter proportional
to $\Delta^{-2}$ \cite{Ben89d,Bar90}, and variation
with the deformation is expected \cite{Mol81}
to result in an $\epsilon$$_2$$^{2}$ dependence.
We have therefore made the following modification:
$\ln F^{calc} \rightarrow  \ln  F^{calc} \times
f(\epsilon,\Delta)$, where
$f(\epsilon,\Delta) =
(\epsilon_2/\epsilon_2^{(0)})
\times (R_{\Delta}/R_{\Delta}^{(0)})^{1/2}$
and $R_{\Delta}$ is an
appropriately chosen function
of $\Delta$$_{p}$ and $\Delta$$_{n}$.  The
value of $\epsilon$$_2^{(0)}$ was fixed at 0.25.
Since the inertial parameter depends on both
pairing gaps, $\Delta$$_{p}$ and $\Delta$$_{n}$,
a functional form for $R_{\Delta}$ must be
chosen.  Motivated by the method used in Ref.
\cite{Ben89d} to calculate the inertial parameter,
the expression
$R_{\Delta} = \sum G_{i} \Delta_{i}^{-2}
/ \sum G_{i}$
was chosen, where the index $i$
refers to neutrons and protons, and the pairing
strengths are $G_{n}$=(18 MeV)/A and $G_{p}$
= (21 MeV)/A \cite{Ben89d}.  (It should be
noted,  however,  that other authors \cite{Bar90}
have proposed somewhat different functional
forms for this dependence on the pair gaps.)
Since the calculations do not show very large
variations in the pair gaps as a function
of $\gamma$,  the pair gaps used in the
tunneling calculations have been determined simply
from the odd-even mass differences \cite{Jen84}.
The quadrupole deformations, $\epsilon_2$, have been
taken from the calculated values at the saddle
point.  The reference value of $R_{\Delta}$
(denoted as  $R_{\Delta}$$^{(0)}$) was fixed at
1.21 MeV$^{-2}$,  the value for $^{174}$Hf.  A
summary of the dynamical quantities  for the
14$^{+}$ states is included in Table VI.

	Before comparing theory and experiment for
the 14$^{+}$ isomers,  the model was tested
on a series of 6$^{+}$ isomers in nuclei in
this region.  Data are available for a large
sample of nuclei, all having 6$^{+}$ isomers
decaying through transitions of the same
E2 multipolarity,
and the decays to the 4$^{+}$ levels
of the (K=0) ground-state bands
represent an excellent
test of the predictive power of the model.
The results of the calculations, shown in
Fig. \ref{tunneling_6plus},
exhibit a clear correlation
between the calculated and observed hindrance
factors.  The correlation between $F^{exp}$
and $F^{calc}$ is not significantly affected,
for example, by the specific choice of
the functional
form of $R_{\Delta}$,  although the slope of
the line is subject to some change, introducing
some ambiguity in extracting a value of D.
Subject to this uncertainty,  the inertial
parameter required to fit the data is D $\sim$
30 MeV$^{-1}\hbar^{2}$rad$^{-2}$,  which is
of the right order
of magnitude compared to a previous estimate
of $\sim$13 MeV$^{-1}\hbar^{2}$rad$^{-2}$ at
zero spin \cite{Mol81}.

	Returning to the K=14 isomers,  the results
for  $^{176}$W and $^{174}$Hf are shown in
Fig. \ref{tunneling_14plus}.
As in the case of the 6$^{+}$
isomers,  the tunneling calculations reproduce
the variations of the hindrance factors.
The extracted inertial parameter is D$\sim$60
MeV$^{-1}\hbar^{2}$rad$^{-2}$.
Remembering that the absolute
hindrance factors span many orders of magnitude,
it is encouraging that data on the 6$^{+}$ and 14$^{+}$
isomers can be reproduced with values of the
$D$ that are of the right order of magnitude
compared to the estimate of Ref. \cite{Mol81}.
It should be re-emphasized that the goal
here was not to extract an accurate value
of $D$,  but simply to investigate the possibility
that systematic variations in the hindrance
factors can be understood using a physically
acceptable value of $D$.  The fitted value of
$D$ is quite sensitive to some of the phenomenological
assumptions,  such as the form of the function
$f(\epsilon,\Delta)$.  Nevertheless,  the
inertial parameter is a quantity of great
physical interest,  and the results presented
here are encouraging as an indication of the
future possibilities for extracting information
about dynamics from this type of data,  provided
that appropriate refinements are included
in the model.  The 6$^{+}$ data
provide support for the broad predictive power of
the model, although these states
follow the ordinary pattern of decay of K-isomers,
and could perhaps be explained in other
frameworks.  More importantly,
the qualitatively new pattern
of decay observed for the 14$^{+}$ isomer
in $^{176}$W provides evidence that a new
mechanism of decay is being observed,  and
that the agreement of the calculations with
experiment is unlikely to be fortuitous.

	An important aspect of the
experimental observations
in $^{176}$W that could not be addressed in
the simple $\gamma$-tunneling model is the complete
absence of decays of the K=14 isomer to the
intermediate-K states.  To address this issue
one needs cranking models where the angular
momentum vector is not restricted to lie along
a principal axis.  An obvious next step is
to expand the $\gamma$-tunneling calculations
to include as a second degree of freedom the
tilting angle $\vartheta = arccos(K/\sqrt{J(J+1)})$,
which is the angle between the
angular momentum vector and the axis of symmetry.
It
appears that $^{176}$W, in which the
ordinary pattern of decay to states of
similar K is absent, is an extreme case,
in which the $\gamma$ degree of freedom
is dominant.
Such a calculation would not only provide
a way to evaluate the relative importance
of the $\gamma$ and $\vartheta$ degrees of freedom
in intermediate cases,  but would also allow
the calculation of the decays to states with
intermediate K,  which are outside the model
space of principal-axis cranking calculations.
 The calculation of cranking wave-functions
at various values of $\vartheta$ is a subject
of much current theoretical work
\cite{Fra82,Fra91,Don89,Don90a,Don90b,Don90c}.

	A schematic method for understanding the
fluctuations in the orientation or ``tilting''
angle $\vartheta$ has
been suggested recently \cite{Fra93b}, in which
a spectrum of states is calculated near the
yrast line, corresponding to a variety of
K-values.  These states are subsequently coupled
together with randomly chosen matrix elements.
The matrix elements are assumed to vary from
zero to
$V_{ij} = V_{0}\cdot f_{V}^{\mid K_{i}-K_{j}\mid}$
,
where V$_{0}$ and $f_{V}$ are phenomenological
parameters to be fitted to the data.  Matrix
elements connecting states with
$\mid K_{i}-K_{j} \mid>1$ are
intended to describe the non-axial shape fluctuations.
The hindrance factors thus calculated are
mainly sensitive to the level densities.

	In order to evaluate the applicability of
this model to the present data,  we have calculated
the spectrum of J=14 states located near
the yrast line in $^{176}$W using the unified
model.  The intrinsic states came from calculations
using the Nilsson model \cite{Ben89e},  with
particle-number projection to avoid the introduction
of spurious particle-particle and hole-hole
states for configurations involving quasiparticles
far from the Fermi level.  A deformation of
$\epsilon$=0.23 was used,  and the pair gaps were adjusted
to reproduce the typical energy differences
among 0-qp, 2-qp and 4-qp states.
The moments of inertia were taken to be $\Im$=36,
46, 56 $\hbar^{2}$ MeV$^{-1}$,  for 0-qp,  2-qp,
and 4-qp states,  respectively.  In contrast
to the calculations of Ref. \cite{Fra93b} where
all states were included, only the states
with the same spin and parity were considered.
Rotational-alignment effects were ignored
in these calculations for simplicity, especially
since the core rotational frequencies involved
are all below the values at which alignments
occur,  and the goal was only to compare the
densities of states.  The results for $^{174}$Hf
and $^{176}$W are shown in Fig.
\ref{level_density}.
The density of 14$^{+}$ states is essentially
the same in both cases.  The Fermi level changes
by only 0.4 MeV between Hf and W because
of the high density of
proton levels.  Since this change in the Fermi
level is significantly less than the pair
gap, $\Delta$,
the quasiparticle energies,
$\sqrt{(E-\lambda)^{2}+\Delta^{2}}$,
change very little.  For the range of energies
shown in the figure,  the average density
of states for a given value of K is 1.1
MeV$^{-1}$
for $^{176}$W,  and
1.3 MeV$^{-1}$ for $^{174}$Hf.
Within the perspective of the model, this result
implies that orientation fluctuations in $^{176}$W
should be slightly smaller than those in $^{174}$Hf,
but the experimental observations are
to the contrary.  The calculations
of Ref. \cite{Fra93b} also result consistently
in a strong correlation between the hindrance
factors and the degrees of K-forbiddenness,
and such a correlation is not observed in
the decay of the isomer in $^{176}$W.

	It has been suggested \cite{Fra93b}
that $\gamma$-tunneling
should actually become less competitive at
high spin,  since the inertial parameter varies
with pairing as $\Delta^{-2}$,  and pairing
is weaker at high spins.  However,
BCS calculations
with particle-number projection, such as
those carried out here,
show only a very slow decrease of $\Delta$.
Our fitted values of the
inertial parameter $D$
seem to exhibit an increasing
trend as a function of spin,
but we do not presently
consider our method of fitting
sufficiently reliable
to constitute an experimental measurement
of $D$ as a function of spin.

\section{CONCLUSIONS}

	Until the last decade,  it appeared that
a satisfactory understanding of the systematics
of  K-violating transitions had been achieved
\cite{Lob68},  with universal hindrance
factors of about 100 per degree of K-forbiddenness.
More recent data \cite{Cho88,Hf174},
including the present work, have now shown
that significant deviations from the previously
established systematics can occur.  In particular,
a K=14 four-quasiparticle isomer was found
in $^{176}$W for which these deviations represent
the only detectable mode of decay;  the usual
decay paths, which tend to minimize the degree
of K-violation, were not observed at all.
The decay of this isomer is even more intriguing
when compared with the decay of another K=14
isomer, with an identical configuration, in
the neighboring nucleus $^{174}$Hf, where
the hindrance factors for similar $\Delta$K=14
transitions are larger by two orders of magnitude.
Simple models of Coriolis
mixing and $\gamma$-tunneling
were applied in an attempt to account for
these observations.  It was found that the
present results on the decay of K=14 isomers
cannot be reproduced with simple extensions
of the accepted picture of Coriolis mixing.
In contrast, a $\gamma$-tunneling framework
seems to provide a natural explanation for
the marked differences between the $^{176}$W
and $^{174}$Hf decays,  although the issue
of the lack of decay of the $^{176}$W isomer
to intermediate-K states falls outside the
model space of these calculations.  The $^{176}$W
nucleus may have provided the first example
of a high-K isomer whose decay is dominated
by large-amplitude, non-axial fluctuations
of the nuclear shape.  In most nuclei, however,
both shape fluctuations and fluctuations
in the orientation of the nucleus with respect
to the angular momentum vector will be important
at moderate spins.  Calculations incorporating
both degrees of freedom are clearly desirable.

\section*{ACKNOWLEDGEMENTS}

	The authors thank J. Greene for the fabrication
of the Nd targets,  T. Bengtsson and the Warsaw
theory group for making their computer codes
available to the physics community,  and S.
Frauendorf for stimulating discussions.

This work was supported in part by U.S. DOE
Contract Nos. DE-FG02-91ER-40609, No.
W-31-109-ENG-38 and DE-FG02-94ER40848.



\begin{figure}
\caption{Level scheme of $^{176}$W deduced
from this work.  Gamma-ray transitions are
shown with their energies in keV, and states
with their spin and
parity.  The labels for the rotational bands
are shown in circles above or below the bands
(see text).}
\label{level_scheme}
\end{figure}

\begin{figure}
\caption{Sample gamma-ray spectra for the
K=0 states: (a) spectrum in coincidence with
the 686 keV 22$^{+}$$\rightarrow$20$^{+}$ transition,
showing the high-spin s band; (b) sum of spectra
in coincidence with the 694, 718 and 767 keV
members of the non-yrast part of the g band.
 Both spectra have the additional coincidence
requirement of BGO fold $\geq$10.}
\label{k0_spectra}
\end{figure}

\begin{figure}
\caption{Gamma-ray spectra based on coincidence
events
in which both Ge detectors registered delayed
gamma rays: (a) total projection;  (b) spectrum
in coincidence with the 596 keV g-band transition;
(c) spectrum in coincidence
with the 408 keV transition from the s band.
All spectra have the additional coincidence
requirement of delayed BGO fold$\geq$1.
Transitions
involved in the decay of the 35-ns
isomer are labeled in keV.  Transitions in
the g band populated by
the decay of the isomer
are marked with triangles.}
\label{dd_spectra}
\end{figure}

\begin{figure}
\caption{Measurement of the half-life
of the 35-ns isomer from the reaction
$^{50}$Ti($^{130}$Te,4n)
with a recoil-shadow geometry (see text).
Three time spectra are shown, based on
background-subtracted slices taken from
a histogram of gamma-ray energy versus
time.  All times were measured relative
to the RF pulse from the accelerator,
which relates to the time of arrival of
the beam.  The lines through the data
all show half-lives of 35 ns, with
amplitudes fitted individually
to the three
spectra.  The calibration of the time
axis can be read directly
from the total projection
at the bottom, which shows the 82-ns
beam-pulsing period of the accelerator.
The data have been renormalized by
factors of 100, 14, 1 and 1/2000,
respectively, for the 714, 491, and
408 keV slices and the total
projection.}
\label{half_life}
\end{figure}

\begin{figure}
\caption{Demonstration of the sensitivity
provided by requiring a coincidence with the
delayed firing of four BGO detectors.  The
transitions from the decay of the 35-ns isomer
are not visible in the ungated spectrum (a),
 but are quite prominent in the spectrum with
the coincidence requirement (b).}
\label{kdel_spectra}
\end{figure}

\begin{figure}
\caption{Excitation energy as a function of
J(J+1) for the g,  s and 0$_{2}{^+}$  bands,
 and the 14$^{+}$ isomer in $^{176}$W.  A
rotational band with perfect J(J+1) level-spacing
would appear on this figure as a straight
line.  An arbitrary reference
value of (9 keV)$\times$J(J+1)
has been subtracted from the excitation energies
in order to highlight small deviations from
a J(J+1) level-spacing,  due either to interactions
between bands or to other effects such as
centrifugal stretching (see text for a detailed
discussion).   The three arrows show the direct
decay paths of the K=14 isomer to K=0 states,
 and the associated intensities are given
as a fraction of the total strength with which
the band is populated.}
\label{ex}
\end{figure}

\begin{figure}
\caption{Systematics of 2-quasiparticle
states observed experimentally in
the W and Hf isotopes.}
\label{syst_2qp}
\end{figure}

\begin{figure}
\caption{Comparison of observed and calculated
2-quasiparticle Routhians in $^{176}$W.
Each band is labeled with the quantum numbers
K$^{\pi}$ or K$^{\pi}$,$\alpha$.}
\label{routhians}
\end{figure}

\begin{figure}
\caption{Comparison of measured (points)
and calculated (line)
B(M1)/B(E2) ratios in band 1, the band built
on the 35-ns isomer in $^{176}$W (see text).}
\label{m1_e2}
\end{figure}

\begin{figure}
\caption{Comparison of the Coriolis mixing
calculations with experiment.  The $x$ axis
shows the minimum calculated hindrance factor,
 under the extreme assumption that 100\% of
the s band wave-function has K=K$_{max}$ (see
text).  The arrow on the s,12$^{+}$ data-point
for $^{176}$W indicates an experimental
upper limit on the corresponding branching
ratio.  The dashed line represents the
assumption that the hindrance factors are
as small as possible in the context of
the model, and thus data-points lying below
this line are unphysical.}
\label{coriolis}
\end{figure}

\begin{figure}
\caption{Schematic representation of the
$\gamma$-tunneling model.  The thick lines depict
the potential energy as a function of $\gamma$,
while the thin lines illustrate the wave-function.}
\label{tunneling_schematic}
\end{figure}

\begin{figure}
\caption{Calculated potential energy curves
for $^{176}$W and $^{174}$Hf as functions
of $\gamma$ deformation for positive parity
and $<$J$_x$$>$=14.}
\label{potential_energy}
\end{figure}

\begin{figure}
\caption{Test of the $\gamma$-tunneling interpretation
as applied to K=6 states in the rare-earth
region.  The dashed line is obtained for a
value of the inertial parameter
D=30 MeV$^{-1}\hbar^{2}$rad$^{-2}$.
The width of each data-point indicates
the range given in the text for
the unhindered strength
S.}
\label{tunneling_6plus}
\end{figure}

\begin{figure}
\caption{Test of $\gamma$-tunneling for the
14$^{+}$ isomers.  The dashed line in this
case is for an inertial parameter
D=60 MeV$^{-1}\hbar^{2}$rad$^{-2}$.
As in Fig. 13, the width of each
data-point indicates
the range given in the text for
the unhindered strength S.
The data-points with arrows
pointing upwards
represent lower limits on
the hindrance factors.}
\label{tunneling_14plus}
\end{figure}

\begin{figure}
\caption{Excited 14$^{+}$ states
calculated in $^{176}$W
and $^{174}$Hf as a function of the tilting
angle.  Note that because of zero-point rotation,
states with K=J have tilting angles slightly
greater than zero.}
\label{level_density}
\end{figure}

\begin{table}
\caption{Energies and intensities
of gamma rays in $^{176}$W observed
in this work, and spin
assignments (from measured DCO ratios) for
the initial and final states of the transitions.}
\begin{tabular}{llll}
$E_{\gamma}$ (keV)	& $I_{\gamma}$ $^{(a)}$ &
$K,J^{\pi}$\hspace{20mm}$\rightarrow$ & $K,J^{\pi}$	\\ \hline
107.8(1)	& 25 $^{(b)}$	& 0$_{g}$,2$^{+}$	& 0$_{g}$,0$^{+}$ \\
174.2(3)	& .6(1)	& 4,6$^{(-)}$	& 4,5$^{(-)}$ \\
186.3(2)	& .7(2)	& (13,14)	& (13,13) \\
202.1(6)	& $\sim$.1	& $^{(c)}$ \\
210.5(1)	& .07$^{(d)}$	& $^{(e)}$ \\
218.0(4)	& .5(2)	& (13,15)	& (13,14) \\
219.0(4)	& .6(2)	& $^{(f)}$ \\
222.8(1)	& 1.1(2)	& 14,15$^{+}$	& 14,14$^{+}$ \\
230.7(1)	& 1.0(2)	& 0$_{s}$,14$^{+}$	& 0$_{g}$,14$^{+}$ \\
238.2(1)	& .9(2) $^{(d)}$	& 14,16$^{+}$	& 14,15$^{+}$ \\
239.7(1)	& 82(6)	& 0$_{g}$,4$^{+}$	& 0$_{g}$,2$^{+}$ \\
247.7(2)	& .6(2)	& (13,16)	& (13,15) \\
251.9(3)	& .5(1)	& 4,8$^{(-)}$	& 4,7$^{(-)}$ \\
256.2(1)	& .6(1) $^{(d)}$	& 14,17$^{+}$	& 14,16$^{+}$ \\
266.6(1)	& .44(8) $^{(d)}$	& $^{(e)}$ \\
267.6(2)	& .8(1)	& 6,8$^{(+)}$	& 6,6$^{(+)}$ \\
272.3(1)	& .6(1)	& 4,7$^{(-)}$	& 4,5$^{(-)}$ \\
273.9(1)	& 3.0(2)	& 4,6$^{(-)}$	& 4,4$^{(-)}$ \\
275.0(2)	& .6(2)	& (13,17)	& (13,16) \\
275.9(1)	& .5(1) $^{(d)}$	& 14,18$^{+}$	& 14,17$^{+}$ \\
292.0(1)	& 1.3(1)	& (7,9)	& (7,7) \\
293.5(1)	& .3(1) $^{(d)}$	& 14,19$^{+}$	& 14,18$^{+}$ \\
297.8(1)	& .20(4)	& $^{(g)}$ \\
300.6(3)	& .4(1)	& 4,10$^{(-)}$	& 4,9$^{(-)}$ \\
301.3(3)	& .5(1)	& (13,18)	& (13,17) \\
308.6(1)	& .3(1) $^{(d)}$	& 14,20$^{+}$	& 14,19$^{+}$ \\
313.7(1)	& .23(5) $^{(d)}$	& $^{(g)}$ \\
315.8(1)	& .28(6) $^{(d)}$	& $^{(e)}$ \\
322.4(2)	& $^{(h)}$	& 14,21$^{+}$	& 14,20$^{+}$ \\
326.0(4)	& .3(1)	& (13,19)	& (13,18) \\
331.7(1)	& .27(5) $^{(d)}$	& $^{(g)}$ \\
334.6(1)	& 3.4(2)	& 4,9$^{(-)}$	& 4,7$^{(-)}$ \\
334.8(4)	& .11(4) $^{(d)}$	& 14,22$^{+}$	& 14,21$^{+}$ \\
339.8(2)	& .6(1)	& 6,10$^{(+)}$	& 6,8$^{(+)}$ \\
347.6(3)	& $^{(h)}$	& $^{(g)}$ \\
348.2(2)	& $^{(h)}$	& 14,23$^{+}$	& 14,22$^{+}$ \\
348.8(1)	& 5.5(3)	& 4,8$^{(-)}$	& 4,6$^{(-)}$ \\
350.8(1)	& 79(2)	& 0$_{g}$,6$^{+}$	& 0$_{g}$,4$^{+}$ \\
354.6(1)	& .8(2) $^{(d)}$	& $^{(e)}$ \\
359.0(3)	& .7(1)	& 4,9$^{(-)}$	& 0$_{g}$,10$^{+}$ \\
362.1(3)	& $^{(h)}$	& $^{(g)}$ \\
363.0(2)	& .5(1)	& 0$_2$,8$^{+}$	& 0$_2$,6$^{+}$ \\
374.0(3)	& .19(4) $^{(d)}$	& $^{(g)}$ \\
374.6(1)	& 2.9(2)	& (7,11)	& (7,9) \\
382.8(1)	& 4.1(3)	& 4,10$^{(-)}$	& 4,8$^{(-)}$ \\
397.0(3)	& .7(2)	& 8,8$^{(-)}$	& 4,6$^{(-)}$ \\
401.6(1)	& 5.5(3)	& 4,11$^{(-)}$	& 4,9$^{(-)}$ \\
405.0(3)	& $<$.1	& (13,15)	& (13,13) \\
405.4(3)	& .7(2)	& (5,7)	& (5,5) \\
408.4(1)	& 2.8(2)	& 0$_{s}$,14$^{+}$	& 0$_{s}$,12$^{+}$ \\
418.0(1)	& 2.0(6)	& 0$_{s}$,12$^{+}$	& 0$_{s}$,12$^{+}$ \\
430.8(2)	& .9(2)	& 0$_2$,10$^{+}$	& 0$_2$,8$^{+}$ \\
431.6(3)	& 1.3(2)	& (5,9)	& (5,7) \\
434.7(1)	& 2.3(2)	& 0$_{s}$,12$^{+}$	& 0$_2$,10$^{+}$ \\
440.4(1)	& 56(2)	& 0$_{g}$,8$^{+}$	& 0$_{g}$,6$^{+}$ \\
441.0(3)	& 1.8(4)	& 8,10$^{(-)}$	& 8,8$^{(-)}$ \\
443.7(1)	& .7(3)	& 6,12$^{(+)}$	& 6,10$^{(+)}$ \\
445(1)	& .25(6) $^{(i,j)}$	& $^{(k)}$ \\
445.5(1)	& 4.4(2)	& 4,12$^{(-)}$	& 4,10$^{(-)}$ \\
446.7(1)	& 3.0(2)	& (7,13)	& (7,11) \\
460.9(1)	& 1.8(4)	& 0$_{s}$,16$^{+}$	& 0$_{s}$,14$^{+}$ \\
461.7(2)	& .27(5) $^{(d)}$	& 14,16$^{+}$	& 14,14$^{+}$ \\
466.5(5)	& .2(2)	& (13,16)	& (13,14) \\
472.1(1)	& 6.0(2)	& 4,13$^{(-)}$	& 4,11$^{(-)}$ \\
473.1(3)	& 1.6(3)	& 8,12$^{(-)}$	& 8,10$^{(-)}$ \\
489.0(3)	& 1.1(3)	& 8,10$^{(-)}$	& 4,8$^{(-)}$ \\
490.9(2)	& .4(1) $^{(i,j)}$	& $^{(k)}$ \\
494.8(3)	& 1.0(2)	& (5,11)	& (5,9) \\
494.9(1)	& .39(5) $^{(d)}$	& 14,17$^{+}$	& 14,15$^{+}$ \\
508.4(1)	& 46(2)	& 0$_{g}$,10$^{+}$	& 0$_{g}$,8$^{+}$ \\
508.8 $^{(l)}$	& 1.8(5)	& 0$_{s}$,18$^{+}$	& 0$_{g}$,16$^{+}$ \\
512.2(3)	& 2.0(4)	& 8,14$^{(-)}$	& 8,12$^{(-)}$ \\
512.2(3)	& .05(3) $^{(i,j)}$	& $^{(k)}$ \\
513.4(1)	& 2.8(5)	& (7,15)	& (7,13) \\
518.6(3)	& .29(6) $^{(i,j)}$	& $^{(k)}$ \\
522.1(1)	& 3.2(2)	& 4,14$^{(-)}$	& 4,12$^{(-)}$ \\
522.5(5)	& .3(2)	& (13,17)	& (13,15) \\
530.5(2)	& .5(3)	& 6,14$^{(+)}$	& 6,12$^{(+)}$ \\
532.5(1)	& .46(6) $^{(d)}$	& 14,18$^{+}$	& 14,16$^{+}$ \\
533.1(1)	& 2.6(4)	& 4,7$^{(-)}$	& 0$_{g}$,8$^{+}$ \\
540.4(1)	& 4.6(3)	& 4,15$^{(-)}$	& 4,13$^{(-)}$ \\
541.1(1)	& 1.0(2)	& 0$_2$,10$^{+}$	& 0$_{g}$,10$^{+}$ \\
545.7(3)	& .6(2)	& (5,13)	& (5,11) \\
553.3(3)	& .8(2)	& 8,16$^{(-)}$	& 8,14$^{(-)}$ \\
557.8(1)	& 28(2)	& 0$_{g}$,12$^{+}$	& 0$_{g}$,10$^{+}$ \\
569.8(1)	& 1.3(3)	& 4,16$^{(-)}$	& 4,14$^{(-)}$ \\
569.8(2)	& .38(5) $^{(d)}$	& 14,19$^{+}$	& 14,17$^{+}$ \\
574.7(1)	& 4.0(6)	& 0$_{s}$,18$^{+}$	& 0$_{g}$,16$^{+}$ \\
576.7(1)	& 2.6(6)	& (7,17)	& (7,15) \\
579.1(5)	& .4(1)	& 8,12$^{(-)}$	& 4,10$^{(-)}$ \\
595.9(1)	& 19(1)	& 0$_{g}$,14$^{+}$	& 0$_{g}$,12$^{+}$ \\
600.7(5)	& .6(2)	& (5,15)	& (5,13) \\
600.8(1)	& 3.1(4)	& 4,17$^{(-)}$	& 4,15$^{(-)}$ \\
603.0(2)	& .37(6) $^{(d)}$	& 14,20$^{+}$	& 14,18$^{+}$ \\
607.7(1)	& 1.0(1)	& 4,18$^{(-)}$	& 4,16$^{(-)}$ \\
611.0(1)	& 3.0(5)	& 0$_{s}$,20$^{+}$	& 0$_{s}$,18$^{+}$ \\
612.0(4)	& .27(5) $^{(d)}$	& $^{(g)}$ \\
618.8(2)	& 1.4(4)	& 0$_2$,8$^{+}$	& 0$_{g}$,8$^{+}$ \\
624.0(4)	& 1.0(4)	& 0$_2$,12$^{+}$	& 0$_{g}$,12$^{+}$ \\
625.3(1)	& 8.8(6)	& 0$_{g}$,16$^{+}$	& 0$_{g}$,14$^{+}$ \\
628.0(5)	& 1.6(3)	& 0$_{g}$,18$^{+}$	& 0$_{s}$,16$^{+}$ \\
629.7(2)	& .7(1)	& 4,20$^{(-)}$	& 4,18$^{(-)}$ \\
631.1(2)	& .29(6) $^{(d)}$	& 14,21$^{+}$	& 14,19$^{+}$ \\
632(1)	& .07(4) $^{(i,j)}$	& $^{(k)}$ \\
633.9(1)	& 1.2(2)	& (7,19)	& (7,17) \\
641.2(2)	& .5(2)	& 0$_2$,12$^{+}$	& 0$_2$,10$^{+}$ \\
647.9(4)	& .18(4) $^{(d)}$	& $^{(g)}$ \\
648.0(2)	& 1.2(2)	& 4,19$^{(-)}$	& 4,17$^{(-)}$ \\
650.0(1)	& .4(1)	& 4,22$^{(-)}$	& 4,20$^{(-)}$ \\
656.3(2)	& .25(6) $^{(i,j)}$	& $^{(k)}$ \\
657.6(3)	& .26(6) $^{(d)}$	& 14,22$^{+}$	& 14,20$^{+}$ \\
674.7(1)	& .7(1)	& (7,21)	& (7,19) \\
680.2(4)	& $^{(h)}$	& $^{(g)}$ \\
682.0(4)	& $^{(h)}$	& 14,23$^{+}$	& 14,21$^{+}$ \\
685.5(2)	& 2.1(7)	& 0$_{s}$,22$^{+}$	& 0$_{s}$,20$^{+}$ \\
685.7(3)	& .4(1)	& 4,21$^{(-)}$	& 4,19$^{(-)}$ \\
691.2(3)	& 3.3(3)	& 0$_{s}$,16$^{+}$	& 0$_{g}$,14$^{+}$ \\
693.7(3)	& 2.7(3)	& 0$_{g}$,18$^{+}$	& 0$_{s}$,16$^{+}$ \\
697.0(2)	& 1.4(2)	& 0$_2$,6$^{+}$	& 0$_{g}$,6$^{+}$ \\
701.4(1)	& 2.2(6)	& 4,5$^{(-)}$	& 0$_{g}$,6$^{+}$ \\
708.6(4)	& .25(6) $^{(d)}$	& 14,24$^{+}$	& 14,22$^{+}$ \\
709.5(4)	& .24(5) $^{(d)}$	& $^{(g)}$ \\
714.1(1)	& .8(1) $^{(i)}$	& 14,14$^{+}$	& 0$_{s}$,14$^{+}$ \\
717.4(4)	& .5(1)	& (7,7)	& 8$^{+}$ \\
718.3(3)	& 1.5(4)	& 0$_{g}$,20$^{+}$	& 0$_{g}$,18$^{+}$ \\
735.0(5)	& $^{(h)}$	& $^{(g)}$ \\
752.1(2)	& 0.4(1)	& 0$_{s}$,24$^{+}$	& 0$_{s}$,22$^{+}$ \\
760.9(1)	& 1.6(3)	& 4,11$^{(-)}$	& 0$_{g}$,10$^{+}$ \\
767.3(4)	& .5(2)	& 0$_{g}$,22$^{+}$	& 0$_{g}$,20$^{+}$ \\
785.4(1)	& 1.1(1)	& 4,8$^{(-)}$	& 0$_{g}$,8$^{+}$ \\
808(1)	& .3(1)	& 0$_{s}$,26$^{+}$	& 0$_{s}$,24$^{+}$ \\
826.6(1)	& .8(1)	& 0$_{s}$,14$^{+}$	& 0$_{g}$,12$^{+}$ \\
866 $^{m}$	& $<$.2	& 14,14$^{+}$	& 4,13$^{(-)}$ \\
867.8(1)	& 3.0(4)	& 4,9$^{(-)}$	& 0$_{g}$,8$^{+}$ \\
876.4(1)	& 5(2)	& 4,6$^{(-)}$	& 0$_{g}$,6$^{+}$ \\
884.5(2)	& .25(6) $^{(i)}$	& $^{(k)}$ \\
916.8(3)	& .31(4) $^{(i)}$	& 14,14$^{+}$	& 0$_2$,12$^{+}$ \\
945.0(2)	& .15(4) $^{(i)}$	& 14,14$^{+}$	& 0$_{g}$,14$^{+}$ \\
953.7(1)	& 2.7(2)	& 4,4$^{(-)}$	& 0$_{g}$,4$^{+}$ \\
957.6(1)	& 1.2(1)	& 6,6$^{(+)}$	& 0$_{g}$,6$^{+}$ \\
973.4(2)	& 2.2(4)	& 4,7$^{(-)}$	& 0$_{g}$,6$^{+}$ \\
1010.0(3)	& 1.2(1)	& 7,9	& 0$_{g}$,8$^{+}$ \\
1047(1)	& .6(3)	& 0$_2$,6$^{+}$	& 0$_{g}$,4$^{+}$ \\
1049.8(3)	& .21(6) $^{(i,j)}$	& 	& 0$_{g}$,12$^{+}$ \\
1067.9(3)	& .16(6) $^{(i,j)}$	& $^{(k)}$ \\
1096(1)	& .08(3) $^{(i,j)}$	& $^{(k)}$ \\
1122 $^{m}$	& $< \sim$.1 $^{(j)}$	& 14,14$^{+}$	& 0$_{s}$,12$^{+}$ \\
1159.5(4)	& .3(1)	& 7,7	& 0$_{g}$,6$^{+}$ \\
1225.9(1)	& 3.2(6)	& 6,8$^{(+)}$	& 0$_{g}$,6$^{+}$ \\
1240.6(3)	& .7(2)	& (5,5)	& 0$_{g}$,4$^{+}$ \\
1296.4(3)	& .9(2)	& (5,7)	& 0$_{g}$,6$^{+}$ \\
1541 $^{m}$	& $<$.06	& 14,14$^{+}$	& 0$_{g}$,12$^{+}$
\end{tabular}
$^{(a)}$ Intensities normalized to total intensity
(gamma plus internal conversion) of the 108 keV
transition (100\%).  All intensities are from
coincidence spectra, with conditions on the number
of detectors firing (at least two Ge detectors
and ten BGO detectors).

$^{(b)}$ Based on theoretical internal conversion
coefficient.

$^{(c)}$ Transition in cascade connecting band
9 to band 8.

$^{(d)}$ Intensity from a coincidence spectrum with
conditions that both Ge detectors fire in-beam
and at least 3 BGO detectors fire out-of-beam,
and normalized using intensity of the 223
keV transition.

$^{(e)}$ Transition in cascade connecting the 10-ns
isomer to the 35-ns isomer.

$^{(f)}$ Decay transition of band 9, not placed
in level scheme.

$^{(g)}$ Transition in band 2 built on 10-ns
isomer.

$^{(h)}$ Not measurable.

$^{(i)}$ Intensity from a coincidence spectrum with
conditions that both Ge detectors as well
as at least 2 BGO detectors fire out-of-beam,
and normalized using the intensity
of the 35-ns isomer.

$^{(j)}$ The intensity listed is the intensity of
population through the 35-ns isomer; the intensity
due to prompt feeding could not be measured.

$^{(k)}$ Transition observed in the decay of
the 35-ns isomer.

$^{(l)}$ Energy from sums and differences of other
gamma-ray energies, and not determined directly
from the centroid of a gamma-ray peak in a spectrum.

$^{(m}$ Unobserved transition. An upper limit on
the intensity was
extracted to allow the calculation of a lower limit
on the corresponding hindrance factor.

\label{Table I}
\end{table}

\begin{table}
\caption{Intensities of transitions associated
with the decay of the 35-ns isomer, listed as
percentages of the total flux
through the isomer (100\%=2.4\% in Table I).}
\begin{tabular}{llll}
$E_{\gamma}$ (keV)	& $I_{\gamma}$  &
$K,J^{\pi}$\hspace{20mm}$\rightarrow$ & $K,J^{\pi}$	\\ \hline
231	& 6(1)	& 0$_s$,14$^{+}$	& 0$_g$,14$^{+}$ \\
240	& 90(17)	& 0$_g$,4$^{+}$	& 0$_g$,2$^{+}$ \\
280	& $<$.1	& 14,14$^{+}$	& 5,13 \\
348	& $<$.2$^{(a)}$	& 14,14$^{+}$	& 8,14$^{(-)}$ \\
351	& 90(17)	& 0$_g$,6$^{+}$	& 0$_g$,4$^{+}$ \\
363	& 4(4)	& 0$_2$,8$^{+}$	& 0$_2$,6$^{+}$ \\
408	& 18(4)	& 0$_s$,14$^{+}$	& 0$_s$,12$^{+}$ \\
418	& 14(3)	& 0$_s$,12$^{+}$	& 0$_g$,12$^{+}$ \\
430	& 9(3)	& 0$_2$,10$^{+}$	& 0$_2$,8$^{+}$ \\
434	& 7(3)	& 0$_s$,12$^{+}$	& 0$_2$,10$^{+}$ \\
440	& 90(4)	& 0$_g$,8$^{+}$	& 0$_g$,6$^{+}$ \\
445	& 10(3) \\
490	& 16(4) \\
508	& 90(4)	& 0$_g$,10$^{+}$	& 0$_g$,8$^{+}$ \\
508	& $<$.2$^{(b)}$	& 14,14$^{+}$	& 6,14(+) \\
513	& 2(1) \\
519	& 12(3) \\
541	& 7(4)	& 0$_2$,10$^{+}$	& 0$_g$,10$^{+}$ \\
558	& 44(4)	& 0$_g$,12$^{+}$	& 0$_g$,10$^{+}$ \\
596	& 7(4)	& 0$_g$,14$^{+}$	& 0$_g$,12$^{+}$ \\
619	& 5(2)	& 0$_2$,8$^{+}$	& 0$_g$,8$^{+}$ \\
624	& 3(2)	& 0$_2$,12$^{+}$	& 0$_g$,12$^{+}$ \\
632	& 3(2) \\
641	& 4(4)	& 0$_2$,12$^{+}$	& 0$_2$,10$^{+}$ \\
656	& 9(3) \\
714	& 33(4)	& 14,14$^{+}$	& 0$_s$,14$^{+}$ \\
775	& $<$.2	& 14,14$^{+}$	& 7,13$^{(-)}$ \\
827	& 5(2)	& 0$_s$,14$^{+}$	& 0$_g$,14$^{+}$ \\
866	& $<$.2	& 14,14$^{+}$	& 4,13$^{(-)}$ \\
884	& 10(3)	&  \\
917	& 13(2)	& 14,14$^{+}$	& 0$_2$,12$^{+}$ \\
945	& 6(3)	& 14,14$^{+}$	& 0$_g$,14$^{+}$ \\
1050	& 7(3) \\
1068	& 5(3) \\
1096	& 3(1) \\
1122	& $< \sim$4	& 14,14$^{+}$	& 0$_s$,12$^{+}$ \\
1541	& $<$2	& 14,14$^{+}$	& 0$_g$,12$^{+}$
\end{tabular}
$^{(a)}$ Intensity was not measurable directly because
of the proximity of the strong 351 keV g-band
transition.  An upper limit was obtained from coincidence
spectra of transitions in band 4 with transitions
in band 1 above the isomer.

$^{(b)}$ Intensity was not measurable directly because
of the proximity of the strong 508 keV g-band transition.
An upper limit was obtained
from coincidence spectra of transitions
in band 8 with transitions in band 1 above
the isomer.
\label{Table II}
\end{table}

\begin{table}
\caption{Comparison of experimental and
calculated branching ratios of out-of-band
to in-band stretched E2 decays in the s band
and g band.  The calculated values are given
for three different interaction matrix elements,
V, showing the quality of the fit as a function
of V.}
\begin{tabular}{lllll}
initial & \multicolumn{4}{c}{B(E2,out)/B(E2,in)} \\
state & exp. & V=25 keV & V=28 keV & V=32 keV \\ \hline
14$^{+}$,s & .008(1) & .002 & .003 & .004 \\
16$^{+}$,s & .24(6) & .11 & .16 & .33 \\
18$^{+}$,s & 1.1(5) & .05 & .07 & .16 \\
18$^{+}$,g & 1.0(3) & .05 & .07 & .16
\end{tabular}
\label{Table III}
\end{table}

\begin{table}
\caption{Configurations assigned to intermediate-K
band-heads, and comparison of experimental
and calculated band-head energies.}
\begin{tabular}{llll}
 & & $E_{x}$ (exp) & $E_{x}$ (theory) \\
K$^{\pi}$ & configuration & (MeV) & (MeV) \\ \hline
4$^{-}$ & $\pi$ 1/2$^{-}$[541]
    $\otimes$  $\pi$ 7/2$^{+}$[404] & 1.301 & 2.1 \\
7$^{-}$	& $\pi$ 9/2$^{-}$[514]
    $\otimes$  $\pi$ 5/2$^{+}$[402] & 1.857 & 2.0 \\
8$^{-}$	& $\pi$ 9/2$^{-}$[514]
    $\otimes$  $\pi$ 7/2$^{+}$[404] & 1.972 & 2.0 \\
6$^{+}$	& $\nu$ 5/2$^{-}$[512]
    $\otimes$  $\nu$ 7/2$^{-}$[514] & 1.656 & 1.6
\end{tabular}
\label{Table IV}
\end{table}

\begin{table}
\caption{Hindrance factors in the nuclei $^{176}$W
and $^{174}$Hf for decays of 14$^{+}$
isomers directly to states with K=0.
As defined in section I, F is the hindrance factor
and f is the hindrance factor per degree of
K-forbiddenness.}
\begin{tabular}{llllllll}
final & & \multicolumn{2}{c}{branching ratio} (\%)
    & \multicolumn{2}{c}F & \multicolumn{2}{c}f \\
state & $\nu$ & $^{176}$W & $^{174}$Hf
    & $^{176}$W & $^{174}$Hf & $^{176}$W
    & $^{174}$Hf \\ \hline
g,14$^{+}$ & 13 & 6(3) & 1.04(6)
    & 2.0$\times$10$^{7}$ & 5.6$\times$10$^{9}$
    & 3.6 & 5.6 \\
g,12$^{+}$ & 12 & $<$2 & .27(12)
    & $>$6.0$\times$10$^{5}$ & 5.0$\times$10$^{8}$
    & $>$3.2 & 5.3 \\
s,14$^{+}$ & 13 & 33(4) & & 8.0$\times$10$^{5}$
& & 3.0  \\
s,12$^{+}$ & 12 & $<$4 & .71(8)
    & $>$1.2$\times$10$^{5}$ & 5.2$\times$10$^{6}$
    & $>$2.9 & (3.6) \\
0$_{2}{^+}$,12$^{+}$ & 12 & 13(2)
    & .59(14) & 8.0$\times$10$^{3}$ & 2.4$\times$10$^{7}$
    & 2.3 & 4.1
\end{tabular}
\label{Table V}
\end{table}

\begin{table}
\caption{Calculated barrier, shape and dynamical
parameters used in the $\gamma$-tunneling model
for the 14$^{+}$ states.  $V_{B}$ is defined
as
$V(\gamma=\gamma_{B})-V(\gamma=-120^{\circ})$,
where $\gamma_{B}$ refers to
the saddle-point.}
\begin{tabular}{llllllll}
 & $V_{B}$  & $\epsilon$$_2$ & $\epsilon$$_2$
    & $\epsilon$$_2$ & $E_{zp}$ \\
 & (MeV) & ($\gamma$=--120$^{\circ}$)
    & ($\gamma=\gamma_{B}$) & ($\gamma$=0) & (MeV)
    & $\epsilon$$_2$/$\epsilon$$_2$(0)
    & ($R_{\Delta}$/$R_{\Delta}$(0))$^{1/2}$ \\ \hline
$^{174}$Hf & 3.4 & .26 & .25 & .26 & 0.53 & 0.96 & 1.00 \\
$^{176}$W & 2.5 & .25 & .23 & .25 & 0.43 & 0.90 & 0.94
\end{tabular}
\label{Table VI}
\end{table}


\newpage
\begin{references}

\vspace{0.2in}


\bibitem{Boh75}	A.\ Bohr and B.M.\ Mottelson,
Nuclear Structure, Vol.II, W.A.\ Benjamin,
1975.
\bibitem{Lob68}K.\ E.\ G.\ L\"{o}bner,
Phys.\ Lett.\ {\bf B}26, 369 (1968).
\bibitem{Cho88}	P.\ Chowdhury et al.,  Nucl.\
Phys.\ {\bf A485}, 136 (1988).
\bibitem{Hf174}P.M.\ Walker et al.,  Phys.\
Rev.\ Lett.\ {\bf 65}, 416 (1990); N.L. Gjorup, P.M.
Walker, G. Sletten, M.A. Bentley, B. Fabricius
and J.F. Sharpey-Schafer,
Nucl. Phys. {\bf A582},
369 (1995).
\bibitem{Wal94a}P.M.\ Walker et al., Nucl.\
Phys.\ {\bf A568}, 397 (1994).
\bibitem{Ben89d} T.\ Bengtsson,  R.A.\ Broglia,
E.\ Vigezzi,  F.\ Barranco,  F.\ D\"{o}nau,
and Jing-ye Zhang,  Phys.\ Rev.\ Lett.\ {\bf 62},
2448 (1989).
\bibitem{Soo91}P.C.\ Sood,  D.M.\ Headly,
and R.K.\ Sheline,  Atomic Data and Nuclear
Data Tables {\bf 47}, 89 (1991).
\bibitem{Cro93b} B.\ Crowell,  P.\ Chowdhury,
D.J.\ Blumenthal,  S.J.\ Freeman,  C.J.\
Lister,  M.P.\ Carpenter,  R.\ Henry,  R.V.F.\
Janssens,  T.L.\ Khoo,  T.\ Lauritsen,  Y.\
Liang,  F.\ Soramel and I.G.\ Bearden,  Phys.\
Rev.\ Lett.\ {\bf 72}, 1164 (1994).
\bibitem{Ye91}Danzhao Ye,  thesis,  Dept.\
of Physics, Univ.\ of Notre Dame, (1991) (unpublished).
\bibitem{Dra78}	G.D.\ Dracoulis, P.M.\ Walker
and A.\ Johnson,  J.\ Phys.\ G {\bf 4}, 713 (1978).
\bibitem{Lee84}I.Y.\ Lee, C.\ Baktash and
J.X.\ Saladin, Phys.\ Rev.\ C {\bf 29}, 837 (1984).
\bibitem{Cro93a} B. Crowell, PhD thesis,
Yale University, 1993 (unpublished).
\bibitem{Cro95}	B.\ Crowell et al., to be
published.
\bibitem{Sle90}G.\ Sletten et al.,  Nucl.\
Phys.\ {\bf A520}, 325c (1990).
\bibitem{Kra88}A.\ Kramer-Flecken, PhD thesis,
 Univ.\ Bonn;  J\"{u}lich preprint J\"{u}l-Spez-458,
Jul 1988.
\bibitem{Kra89}A.\ Kramer-Flecken, T.\ Morek,
 R.M.\ Lieder,  W.\ Gast,  G.\ Hebbinghaus,
H.M.\ J\"{a}ger and W.\ Urban,  Nucl.\ Inst.\
Meth.\ {\bf A275}, 333 (1989).
\bibitem{Kho73a}T.\ L.\ Khoo,   F.M.\ Bernthal,
 J.S.\ Boyno and R.A.\ Warner,  Phys.\ Rev.\
Lett.\ {\bf 31}, 1146 (1973).
\bibitem{Wal91}P.M.\ Walker,  G.D.\ Dracoulis,
 A.P.\ Byrne,  B.\ Fabricius,  T.\ Kib\'{e}di
and A.E.\ Stuchberry,  Phys.\ Rev.\ Lett.\
{\bf 67}, 433 (1991).
\bibitem{Ben86b}R.\ Bengtsson,  proc.\ of
the XVIII Mikolajki Summer School,  Mikolajki,
Poland,  1986,  ed.\ Z.\ Wilhelmi and G.\
Szeflinska, Harwood Academic, p. 20.
\bibitem{Ben89c}T.\ Bengtsson,  Nucl.\ Phys.\
{\bf A520}, 201c (1990).
\bibitem{Ben91}	R.\ Bengtsson,  in Proc.\
of the International Conference on High-Spin
Physics and Gamma-Soft Nuclei,  Pittsburgh,
 eds.\ J.X.\ Saladin, R.A.\ Sorensen and C.M.\
Vincent,  World Scientific 1991, p. 289.
\bibitem{Ben89e} T.\ Bengtsson, Nucl.\ Phys.\
{\bf A496}, 56 (1989) and references therein.
\bibitem{Ber76}	F.\ M.\ Bernthal,  C.L.\ Dors,
 B.D.\ Jeltema,  T.L.\ Khoo and R.A.\ Warner,
 Phys.\ Lett.\ {\bf B64}, 147 (1976).
\bibitem{Sun94}Y.\ Sun, S.\ Wen and D.H.\
Feng, Phys.\ Rev.\ Lett.\ {\bf 72}, 3483 (1994).
\bibitem{Wal94b}P.M.Walker and G.D. Dracoulis,
Phys. Rev. Lett. {\bf 72}, 3736 (1994)
\bibitem{Cro94} B. Crowell
et al., Phys. Rev. Lett. {\bf 72}, 3737 (1994).
\bibitem{Fra93a}S. Frauendorf, Nucl. Phys.
{\bf A557}, 259c (1993).
\bibitem{Ber77}	Eh.\ E.\ Berlovich, P.P.\
Vajshnis, V.D.\ Vitman, R.V.\ Moroz andV.K.\
Tarasov, Probl.\ Yad.\ Fid.\ Kosm.\ Luchei
{\bf 7}, 15 (1977).
\bibitem{Gou70}P.F.A.\ Goudsmit, J.\ Konijn
and F.W.N.\ deBoer, Nucl.\ Phys.\ {\bf A151}, 513
(1970).
\bibitem{And74}	H.R.\ Andrews et al., Nucl.\
Phys.\ {\bf A219}, 141 (1974).
\bibitem{Lie74}R.M.\ Lieder et al., Phys.\
Lett.\ {\bf B49}, 161 (1974).
\bibitem{Mik67}O.\ Mikoshiba, R.K.\ Sheline,
T.\ Udagawa and S.\ Yoshida,  Nucl.\ Phys.\
{\bf A101}, 202 (1967).
\bibitem{Bes66}	D.R.\ Bes and R.A.\ Broglia,
 Nucl.\ Phys.\ {\bf 80}, 289 (1966).
\bibitem{Bir75}	B.L.\ Birbair, N.A.\ Voinova
and N.S.\ Smirnova,
Nucl.\ Phys.\ {\bf A251}, 169
(1975).
\bibitem{Bur88} D.G.\ Burke, G.\ Lovhoiden
and T.F.\ Thorsteinsen,
Nucl.\ Phys.\ {\bf A483},
221 (1988).
\bibitem{Cwi87}	S.\ Cwiok,  J.\ Dudek,  W.\
Nazarewicz,  J.\ Skalski and T.\ Werner,
Comp.\ Phys.\ Comm.\ {\bf 46}, 379 (1987).
\bibitem{Ben85}	T.\ Bengtsson and I.\ Ragnarsson,
 Nucl.\ Phys.\ {\bf A436}, 14 (1985).
\bibitem{Don83}	F.\ D\"{o}nau and S.\
Frauendorf, in Proc. Int. Conf on
High Angular Momentum Properties
of Nuclei, Oak Ridge, 1983,
ed. N.\ R.\ Johnson,
Harwood Academic, p.143.
\bibitem{Bar90}	F.\ Barranco, G.F.\ Bertsch,
 R.A.\ Broglia and E.\ Vigezzi,   Nucl.\ Phys.\
{\bf A512}, 253 (1990).
\bibitem{Mol81}P.\ M\"{o}ller and J.R.\
Nix, Nucl.\ Phys.\ {\bf A}361, 117 (1981).
\bibitem{Jen84}A.S.\ Jensen,  P.G.\ Hansen,
 and B.\ Jonson,  Nucl.\ Phys.\ {\bf A}431, 393
(1984).
\bibitem{Fra82}	S.\ Frauendorf,  in Contemporary
Research Topics in Nuclear Physics, ed.\ D.W.\
Feng, M.\ Vallieres,  M.W.\ Guidry and L.L.\
Riedinger, Plenum, New York, 1982, p. 1.
\bibitem{Fra91}	S.\ Frauendorf and T.\ Bengtsson,
Workshop-Symposium on Future Directions in
Nuclear Physics with 4$\pi$ Gamma Detection Systems
of the New Generation,  March 1991,  Strasbourg,
eds. J. Dudek an B. Haas, American Inst. of
Physics.
\bibitem{Don89}F.\ D\"{o}nau, Nucl.\
Phys.\ {\bf A}496, 333 (1989).
\bibitem{Don90a}F.\ D\"{o}nau, Nucl.\
Phys.\ {\bf A}520, 437c (1990).
\bibitem{Don90b}F.\ D\"{o}nau, Nucl.\
Phys.\ {\bf A}517, 125 (1990).
\bibitem{Don90c}F.\ D\"{o}nau,  in Proceedings
of the International Conference on High-Spin
Physics and Gamma-Soft Nuclei,  Pittsburgh,
Sept.\ 1990,  eds.\ J.X.\ Saladin,
R.A.\ Sorensen and C.M.\
Vincent,  World Scientific 1991, p. \ 66.
\bibitem{Fra93b} S.\ Frauendorf,  in Proc.
Int.\ Conf.\ on the Future of
Nuclear Spectroscopy,
Crete,  June 1993, eds.
W. Gelletly, C. A. Kalfas and R. Vlastou,
Institute of Nuclear Physics,
National Center for
Scientific Research Demokritos,
Athens, Greece, 1994.
\end{references}
\end{document}